\documentclass[%
 reprint,
superscriptaddress,
 amsmath,amssymb,
 aps,
]{revtex4-2}

\usepackage{graphicx}
\usepackage{dcolumn}
\usepackage{bm}
\usepackage{hyperref}
\usepackage{siunitx}

\makeatletter
\def\@affil@script#1#2#3#4{%
 \@ifnum{#1=\z@}{}{%
  \begingroup
   \frontmatter@affiliationfont
   \@ifnum{\c@affil<\affil@cutoff}{}{%
    \def\@thefnmark{#1}\@makefnmark
   }%
   \ignorespaces#3%
   \@if@empty{#4}{}{\frontmatter@footnote{#4}}%
  \endgroup
 }%
}%
\makeatother

\begin{document}


\title{Extended depth-of-field light-sheet microscopy\\improves imaging of large volumes at high numerical aperture}

\author{Kevin Keomanee-Dizon}
 \altaffiliation[Present address: ]{Joseph Henry Laboratories of Physics, Princeton University, Princeton, NJ 08544}
\affiliation{%
Translational Imaging Center,
}%

\author{Matt Jones}
\affiliation{%
Translational Imaging Center,
}%

\author{Peter Luu}
\affiliation{%
Translational Imaging Center,
}%
\affiliation{%
Molecular and Computational Biology, University of Southern California, Los Angeles, CA 90089
}%

\author{Scott E. Fraser}
\affiliation{%
Translational Imaging Center,
}%
\affiliation{%
Molecular and Computational Biology, University of Southern California, Los Angeles, CA 90089
}%

\author{Thai V. Truong}%
 \email{tvtruong@usc.edu}
\affiliation{%
Translational Imaging Center,
}%
\affiliation{%
Molecular and Computational Biology, University of Southern California, Los Angeles, CA 90089
}%


\begin{abstract}
  Light-sheet microscopes must compromise between field of view, optical sectioning, resolution, and detection efficiency. High-numerical-aperture (NA) detection objective lenses provide high resolution but their narrow depth of field fails to capture effectively the fluorescence signal generated by the illumination light sheets, in imaging large volumes. Here, we present ExD-SPIM (extended depth-of-field selective-plane illumination microscopy), an improved light-sheet microscopy strategy that solves this limitation by extending the depth of field (DOF) of high-NA detection objectives to match the thickness of the illumination light sheet. This extension of the DOF uses a phase mask to axially stretch the point-spread function of the objective lens while largely preserving lateral resolution. This matching of the detection DOF to the illumination-sheet thickness increases total fluorescence collection, reduces background, and improves the overall signal-to-noise ratio (SNR). We demonstrate, through numerical simulations and imaging of bead phantoms as well as living animals, that ExD-SPIM increases the SNR by more than three-fold and dramatically reduces the rate of fluorescence photobleaching, when compared to a low-NA system with an equivalent depth of field. Compared to conventional high-NA detection, ExD-SPIM improves the signal sensitivity and volumetric coverage of whole-brain activity imaging, increasing the number of detected neurons by over a third.
\end{abstract}

\maketitle


Studies of biological systems, ranging from molecular and cellular to collections of cells and whole organisms, reveals the need to capture spatial and temporal dynamics of multiple components over time.  Optical imaging is well matched to these demands, in basic science, in clinical research, and, in the diagnosis and treatment of human disease.  The optimal imaging of biological dynamics requires that the relevant spatial, temporal, and energetic scales be captured in their natural settings as they take place.  Selective-plane illumination microscopy (SPIM; light-sheet microscopy) is well suited to this challenge, as its unique strategy in selective illumination of only the focal plane enables high contrast, fast, volumetric image acquisition, with lower light exposure than more typical imaging tools such as confocal laser scanning microscopy \cite{Power_Huisken}.  Since its re-introduction close to two decades ago \cite{Siedentopf_Zsigmondy,Huisken_2004}, SPIM has seen many advancements, and has been widely applied to problems ranging from chemistry to developmental biology and neuroscience \cite{Power_Huisken}.

Despite the amazing capabilities of SPIM, its success has motivated further development to meet the challenges of ever more demanding biological specimens, especially those that are too fast, large, noisy, and light sensitive. The excitation of fluorescent labels in the specimen creates fundamental limits on temporal and spatial resolution.  Even an ideal detector with infinite speed and zero noise must wait for enough photons to form an image with acceptable signal-to-noise ratio (SNR) \cite{Pawley,Supatto_2011}. Thus, fundamental to the performance of any microscope is the ability to collect as many of the signal photons as possible. Given the isotropic nature of fluorescence emission, a high-numerical-aperture (NA) detection objective that captures the largest possible emission solid angle yields ideal signal-collection efficiency and resolution. However, a high-NA detection objective necessarily comes with a tight point-spread function (PSF), whose narrow depth of field (DOF) ($\sim\!\!1$-2 \si{\micro\meter} full-width at half-maximum; FWHM) is often a poor match to the thickness of the excitation light sheet, especially in imaging large samples.

In generating the excitation light sheet with an illumination objective, there is a tradeoff between the thickness and the Rayleigh range of the focused light \cite{Power_Huisken}. High-NA illumination produces thinner sheets that can match the narrow DOF of high-NA detection objectives, but with a smaller Rayleigh length and thus smaller field of view (FOV), which is inadequate for imaging large samples. Propagation-invariant beams, such as Bessel-Gauss \cite{Garces-Chavez_2002,Fahrbach_2010,Planchon_2011}, Airy-Gauss beams \cite{Vettenburg_2015,Hosny_2020},
or bound two-dimensional (2D) optical lattices partially improve this tradeoff, by providing a thin ($\sim\!\!1$ \si{\micro\meter} FWHM) light-sheet profile over a $\sim\!\!50$-\si{\micro\meter} FOV \cite{Chen_2014,Gao_2019}.
However, to image dynamic systems having volumetric dimensions of hundreds of microns or more with cellular resolution \cite{Keller_2008,Ahrens_2013,Wolf_2015,Keomanee-Dizon_2020}, low-NA illumination is needed to produce light sheets that span a sufficinetly large FOV. But low-NA illumination also produces thicker light sheets ($\sim\!\!4$-5 \si{\micro\meter} or more), which excites fluorescence outside the narrow DOF of the high-NA detection objective. This mismatch results in not only loss of fluorescence photons but also inclusion of more out-of-focus background, degrading both light-collection efficiency and SNR.

\begin{figure*}
\includegraphics[scale=0.387]{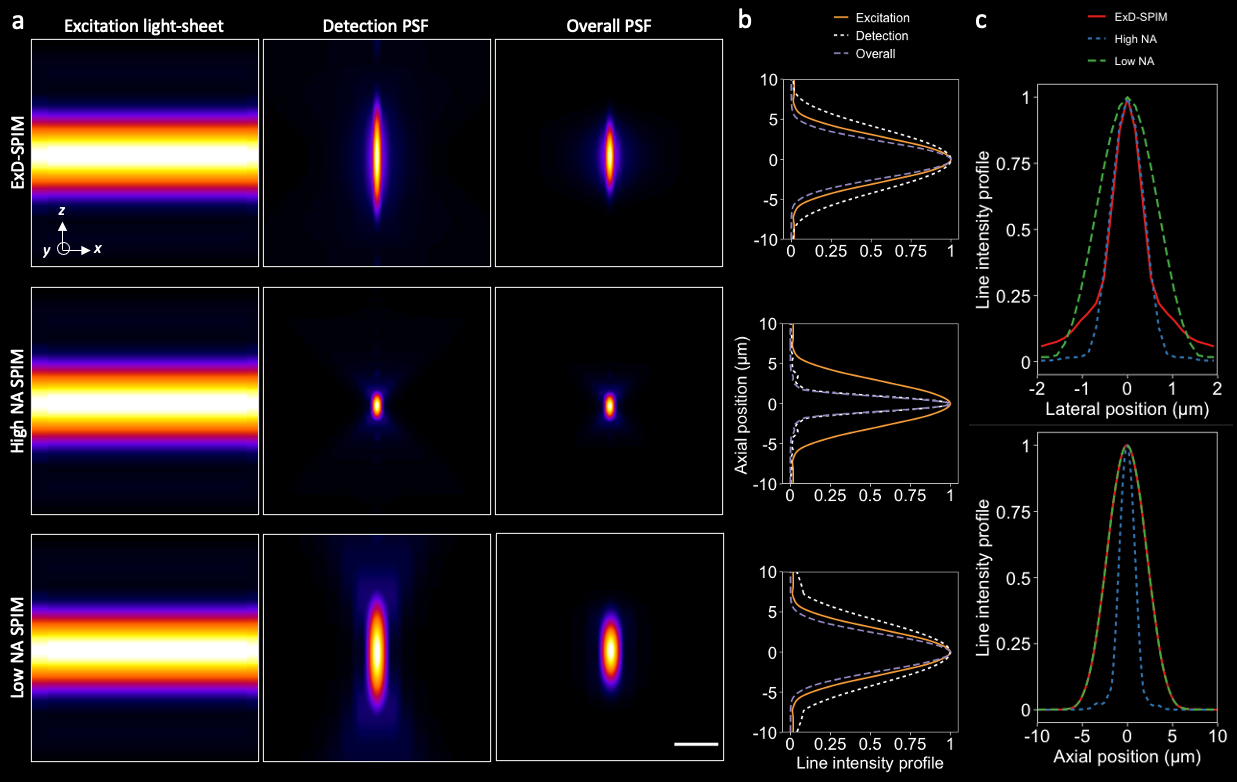}
\caption{\label{fig:ExD}Instantaneous extended depth-of-field light-sheet microscopy at high numerical aperture.\\
\textbf{(a)} Simulated 6-\si{\micro\meter} thick light sheets (left column) and detection PSFs (middle column) for ExD-SPIM (top row; $\mathit{NA}=0.8$), conventional high-NA SPIM (middle row; $\mathit{NA}=0.8$), and low-NA SPIM (bottom row; $\mathit{NA}=0.41$). Overall PSFs (right column) are calculated by multiplying the excitation PSF with the detection PSF. Scale bar, 5 \si{\micro\meter}.\\
\textbf{(b)} $x$-$z$ cross-sectional intensity profiles of the excitation, detection, and overall PSFs. High-NA SPIM (middle row) provides maximal axial confinement of the overall PSF. As a result, however, only a fraction of the excitation light sheet results in useful signal. ExD-SPIM (top row) captures the entire light-sheet thickness by instantaneously extending the DOF of the high-NA detection objective lens, just like a low-NA system (bottom row), but without massively degrading the lateral resolution (c).\\
\textbf{(c)} Lateral (top) and axial (bottom) line intensity profiles of the overall PSFs of the modalities shown in (a).
}
\end{figure*}

To strike a better compromise between light sheet excitation and detection, we have developed ExD-SPIM (extended depth-of-field SPIM), an approach that stretches the PSF of the detection objective in $z$, creating a greater DOF with minimal loss in $x$-$y$ resolution (Fig.~\ref{fig:ExD}). The detection objective DOF is extended to match the light-sheet thickness, exploiting the full benefits of plane illumination: maintaining optical sectioning, reducing background, and achieving greater photon utilization efficiency. The needed DOF extension is achieved through a simple modification of the detection path, adding a ``layer-cake” phase mask \cite{Abrahamsson_2006} conjugate to the pupil plane of the high-NA detection objective in a standard SPIM setup \cite{Madaan_2021} (Fig.~\ref{fig:ExD_optics} and Methods Section~\ref{sec:ExD_optics}). The layer-cake phase mask divides the full pupil of the detection objective into multiple sub-apertures, breaking the coherence of the fluorescence waves traveling to the camera, and creating an axially elongated PSF through the incoherent superposition of the multiple different foci (Methods Section~\ref{sec:ExD_sim_methods}
and Fig.~\ref{fig:layercake}a). Our method simply extends the DOF by several fold (Fig.~\ref{fig:ExD}a and c), sacrificing little lateral resolution (Fig.~\ref{fig:ExD}c) and requiring no post-processing—images can be read out and interpreted directly.

\begin{figure*}
\includegraphics[scale=0.41]{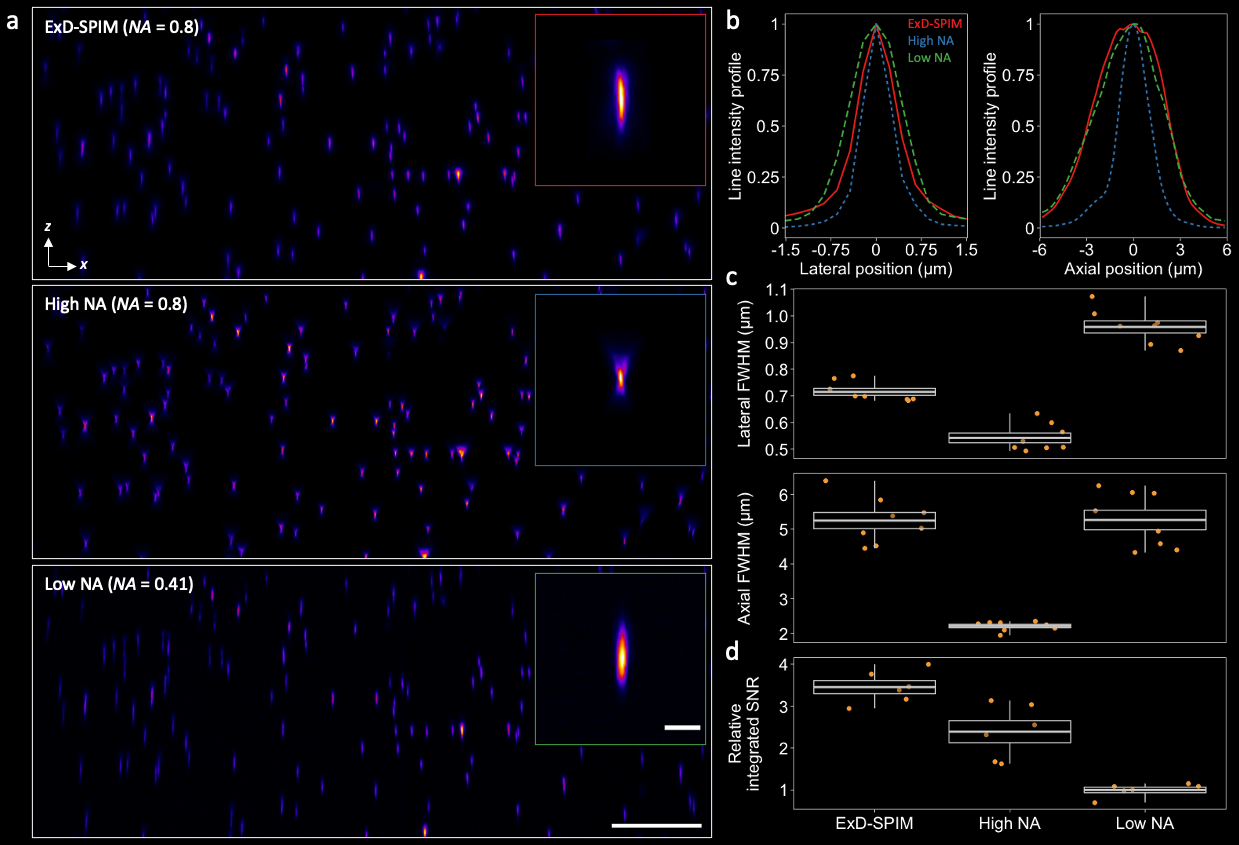}
\caption{\label{fig:ExD_beads}Experimental demonstration of improving light-sheet collection with instantaneous depth-of-field extension.\\
  \textbf{(a)} $x$-$z$ maximum intensity projection of a $375\times555\times150$ ($x$-$y$-$z$) \si{\micro\meter}$^{3}$ beads field captured with ExD-SPIM (top row), high-NA SPIM (middle row), and low-NA SPIM (bottom row). Inset: Representative PSFs computed by averaging over four beads for each modality. Gamma was adjusted (to 0.75) for all modes. Scale bar, 50 \si{\micro\meter} and (inset) 5 \si{\micro\meter}.\\
  \textbf{(b)} Lateral (left) and axial (right) line intensity profiles through the PSFs in the insets of (a). ExD-SPIM shows an axial extent equivalent to the low-NA system with slight lateral blurring.\\
 \textbf{(c)} Mean lateral (top) and axial (bottom) FWHM measurements; for each modality, the same ($N = 8$) beads were chosen from (a). The mean lateral and axial FWHM $\pm$ SD values are ExD-SPIM, 0.71 $\pm$ 0.04 \si{\micro\meter} and 5.25 $\pm$ 0.66 \si{\micro\meter}, respectively; high NA, 0.54 $\pm$ 0.05 \si{\micro\meter} and 2.21 $\pm$ 0.14 \si{\micro\meter}, respectively; low NA, 0.96 $\pm$ 0.06 \si{\micro\meter} and 5.26 $\pm$ 0.8 \si{\micro\meter}, respectively.\\
 \textbf{(d)} Quantification of the integrated SNR (along both the $x$ and $z$ directions; see Methods Section~\ref{sec:SNR}) across multiple beads from a $x$-$z$ summed-intensity projection of (a). ExD-SPIM shows $\sim\!\!3.5\times$ enhancement compared to low-NA SPIM, and a $\sim\!\!45\%$ enhancement over high-NA SPIM, in the integrated SNR.\\
 All boxes are standard deviations; center values are means; whiskers represent the spread of the data.
}
\end{figure*}

ExD-SPIM is simpler, faster and more robust than other SPIM techniques that elongate the DOF by hundreds of microns \cite{Olarte_2015,Tomer_sped,Quirin_2016}.  DOF extension through wavefront coding \cite{Dowski,Olarte_2015,Quirin_2016} requires computational transformation and deconvolution, and loses in signal sensitivity at higher spatial frequencies.  This makes dim features even dimmer and thus nonoptimal in low SNR regimes. Extending the DOF by introducing spherical aberration through the detection objective \cite{Tomer_sped} is only practical at low NA, which limits spatial resolution and light collection.  Increasing the DOF with an active optical device (electrically tunable lens \cite{Fahrbach_2013}, mirror galvanometer \cite{Botcherby_2007}, or deformable mirror \cite{Shain}) by remotely scanning the focus over an extended axial range during a camera exposure yields a time-averaged effective extended DOF, but at low duty cycle (given fluorescence lifetimes are in the nanosecond range, and the fastest devices are currently over an order of magnitude slower). These axial scanning approaches capture signal with far less efficiency and are limited in temporal resolution, compared to the instantaneous capture of photons over the extended axial range of ExD-SPIM.

Numerical simulations of ExD-SPIM validates the performance of our approach.  We compared ExD-SPIM ($\mathit{NA}=0.8$) to conventional SPIM at high NA ($\mathit{NA}=0.8$) and SPIM at low NA ($\mathit{NA}=0.41$). The conventional high-NA mode serves as the high-resolution, diffraction-limited reference, while the low-NA mode serves as the low-resolution, DOF-equivalent reference.  The PSFs were computed using the Debye approximation, as described by Chen \textit{et al}. \cite{ChenChakraborty_2020}, with the overall PSF of each case calculated as the product of the detection PSF and a Gaussian-beam light-sheet with 6-\si{\micro\meter} FWHM thickness (see Methods Section~\ref{sec:ExD_sim_methods}). We chose a 4-layer-cake phase mask to capture the entire light-sheet thickness across the field of view, so that all illuminated fluorophores are instantly in focus and their emitted photons captured with high contrast (Fig.~\ref{fig:ExD}b and d).

We experimentally benchmarked ExD-SPIM performance by measuring the PSF with sub-diffractive fluorescent beads embedded in a mixture of 1.5\% agarose and 10\% Iodixanol \cite{Boothe_2017}. Compared with the high-NA reference, ExD-SPIM lost only $\sim\!\!30\%$ in average lateral resolution across the imaged beads volume; whereas, the low-NA system achieved significantly worse lateral resolution (Fig.~\ref{fig:ExD_beads}a and b). ExD-SPIM extends the PSF by $\sim\!\!2.4$-fold axially, compared to the high-NA system, comparable to the low-NA case (Fig.~\ref{fig:ExD_beads}a and b). The measurements of axial and lateral resolution are in good agreement with theoretical simulations (Fig.~\ref{fig:ExD}b and c). Importantly, we see a significant increase in the integrated SNR (Methods Section~\ref{sec:SNR}) of beads measured over a
$\sim375\times555\times150$ \si{\micro\meter}$^3$
volume: ExD-SPIM SNR = 3.45 $\pm$ 0.38 (mean $\pm$ SD); conventional high-NA SPIM SNR = 2.39 $\pm$ 0.65 (ExD-SPIM $\sim\!\!45\%$ better); low-NA SPIM SNR = 1 $\pm$ 0.16 (ExD-SPIM $\sim\!\!350\%$ better; Fig.~\ref{fig:ExD_beads}d).

The larger DOF of ExD-SPIM compared to the smaller DOF of conventional high-NA SPIM leads to a direct benefit in volumetric imaging, allowing a sparser axial sampling interval in ExD-SPIM. To image the same 3D volume without missing features, fewer $z$-slices are required by ExD-SPIM, as demonstrated by recording a volume of fluorescent beads (Fig.~\ref{fig:undersampling}). This improves the ability of ExD-SPIM to capture specimens with faster dynamics over larger volumes, and reduces the total laser illumination, resulting in less photodamage.

The higher SNR of ExD-SPIM for each 2D optical section permits imaging using lower laser excitation, resulting in lower photodamage. We experimentally validated this by comparing rates of photobleaching during the recording of a single-plane of the fluorescent signals from GFP-labeled vasculature in live larval zebrafish (4-5 days post-fertilization, dpf). For the comparison, the laser power was adjusted to achieve similar SNR in the first image of the time series; exposure times and other imaging parameters were identical over the different samples (Methods Section~\ref{sec:ExD_sample}). ExD-SPIM performed the best: after 4 hours of continuous imaging (70,760 images), the cumulative bleaching loss was only $\sim\!\!5\%$ of the initial fluorescence (Fig.~\ref{fig:photobleach}). The bleaching loss of low-NA SPIM was $\sim\!\!20\%$, $\sim\!\!4$-fold faster (Fig.~\ref{fig:photobleach}). This demonstrates that ExD-SPIM achieves the twin goals of minimizing excitation light exposure and maximizing fluorescence collection for long-term \emph{in vivo} imaging.

\begin{figure*}
\includegraphics[scale=0.52]{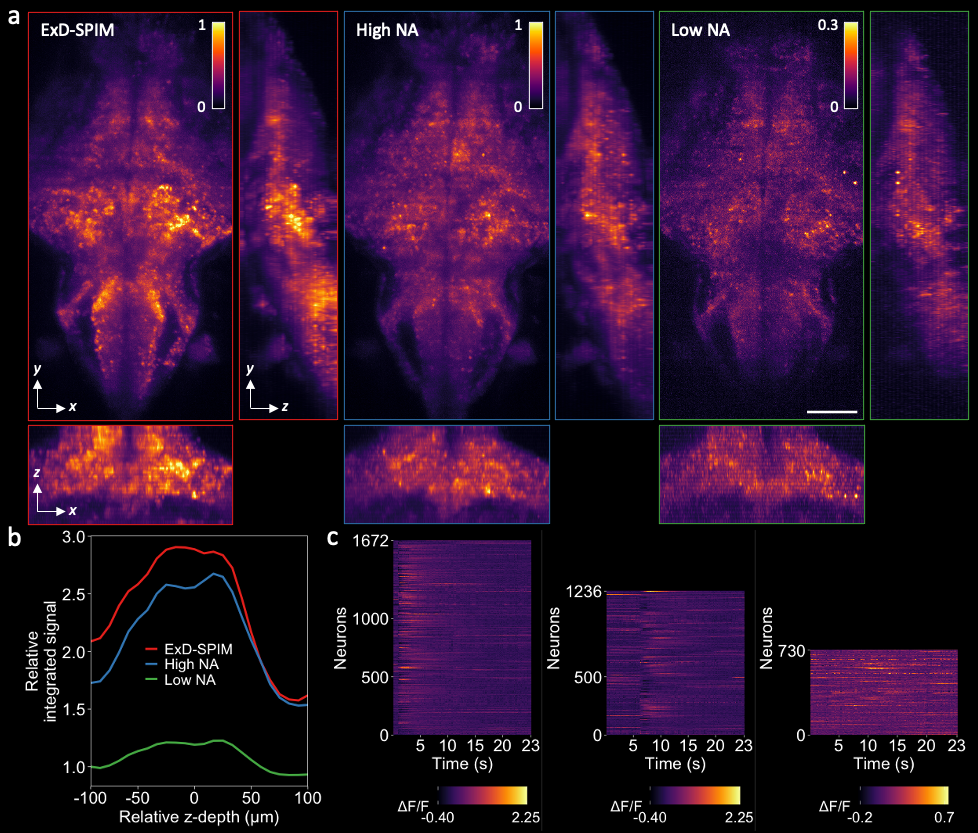}
\caption{\label{fig:ExD_neuro}ExD-SPIM improves whole-brain activity imaging.\\
\textbf{(a)} $x$-$y$, $y$-$z$ (right), and $x$-$z$ (bottom) maximum intensity projections (MIPs) of neural activity in transgenic larval zebrafish (\textit{Tg(elavl3:H2b-GCaMP6s})) at 5-dpf, captured by (left to right) ExD-SPIM (red), high-NA SPIM (blue), and low-NA SPIM (green). Imaging performed at 5.64 Hz of the same animal (Methods Section~\ref{sec:ExD_sample}). Projections show the increased signal collection of ExD-SPIM compared to high-NA and low-NA SPIM. Scale bar, 100 \si{\micro\meter}. Also see Video~\ref{vid:neuro_clip}.\\
\textbf{(b)} Plot shows the total summed intensity of the standard deviation projection (same volume as in (a)) over a 5-minute time series (1692 time points), as a function of $z$-depth. The total integrated signal intensity is improved across the entire image volume with ExD-SPIM compared to SPIM at high NA and low NA.\\
\textbf{(c)} Single-neuron activity traces captured by the different modalities, extracted over a 23-s window from the 4D time-series data of (b), indicating more neurons detected with ExD-SPIM than conventional SPIM at both high and low NA. $\Delta$F/F computed as (F-F\textsubscript{0})/F\textsubscript{0}, where F is the mean fluorescence intensity and F\textsubscript{0} is the bottom eighth percentile of fluorescence over time.
}
\end{figure*}

To test ExD-SPIM on a dynamic volumetric imaging challenge, we imaged whole-brain activity in a 5-dpf transgenic larval zebrafish expressing the calcium indicator GCaMP6s.  ExD-SPIM allowed us to collect with high SNR an image stack of 25 optical sections, spanning 200-\si{\micro\meter} in depth, recorded at 5.64 Hz (Fig.~\ref{fig:ExD_neuro}a). This volumetric imaging rate of $\sim\!\!10^5$ voxels/s was limited only by the readout speed of the detector, not the available signal (Video~\ref{vid:neuro_clip}).
To quantify the integrated signal and determine the number of active cells successfully captured in the time series, we created a 3D volumetric map of the temporal standard deviation for each $x$-$y$ image plane of the 4D data (a 5-minute recording represents 1692 time points). This 3D map highlights the positions of the active neurons, and allowed us to extract single-neuron temporal traces \cite{Madaan_2021,Truong_2020}.
ExD-SPIM provided an overall increase in total integrated signal at every acquired $x$-$y$ image slice, by $\sim\!\!11\%$ compared to high-NA SPIM, and $\sim\!\!220\%$ compared to low-NA SPIM (averaged over the 25 optical sections; Fig.~\ref{fig:ExD_neuro}b). More important, ExD-SPIM detected $\sim\!\!35\%$ more active neurons in the volumetric time series, collecting several hundreds of intracellular calcium transients that otherwise would have gone undetected with conventional high-NA detection (Fig.~\ref{fig:ExD_neuro}c). These results highlight the improvements offered by ExD-SPIM for cellular resolution recording in challenging preparations.

\begin{video}
\href{https://kdizon.github.io/public/ExD_fish-brain.mp4}{\includegraphics[scale=0.425]{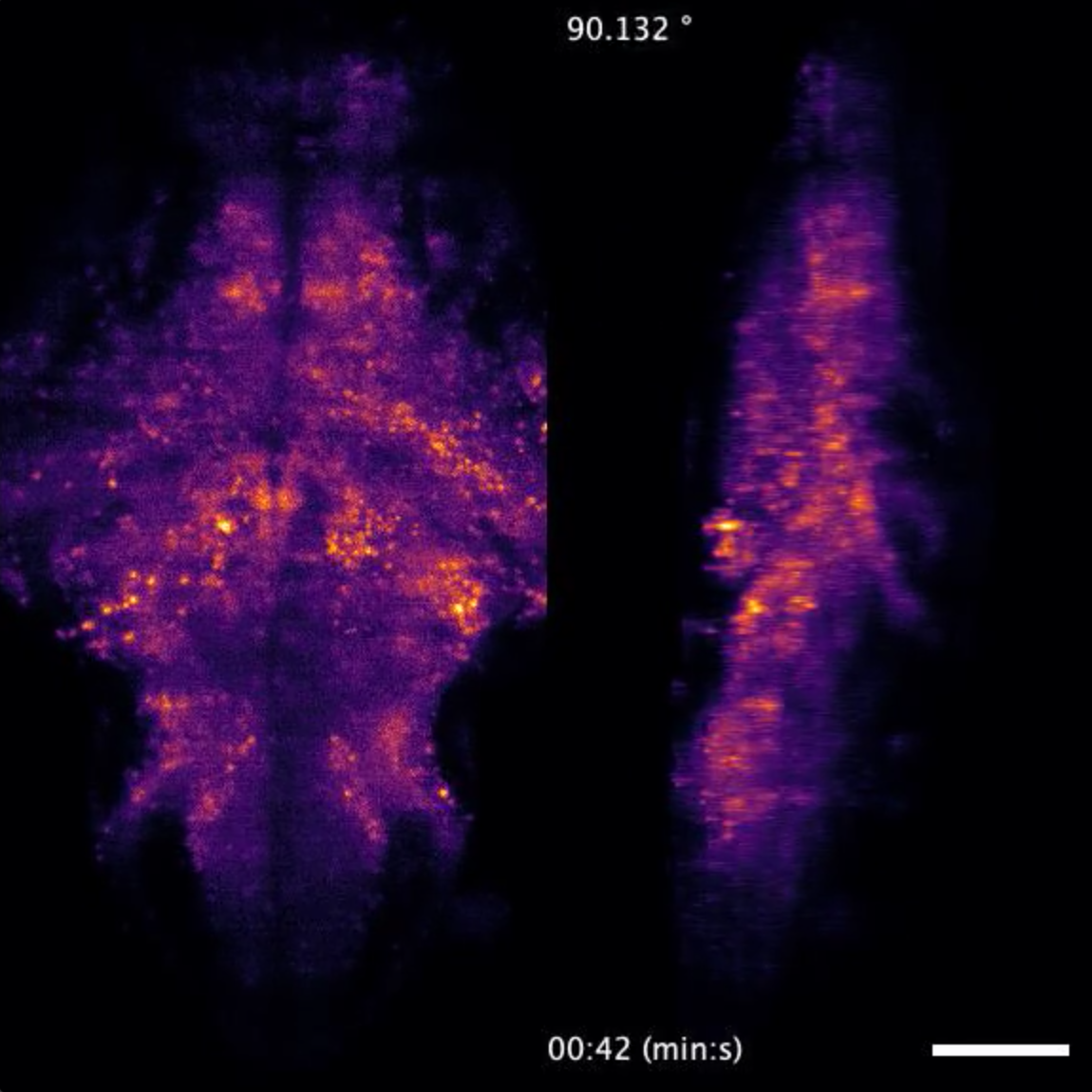}}%
 \setfloatlink{https://kdizon.github.io/public/ExD_fish-brain.mp4}%
 \caption{\label{vid:neuro_clip}%
  ExD-SPIM recording of whole-brain neural activity.\\
Dorsoventral (left) and rotating (right) maximum-intensity projections of a time-series recording of the whole-brain of a 5-dpf transgenic larval zebrafish. Whole-brain functional light-sheet imaging was performed at a volumetric rate of 5.64 Hz. The video is part of the data presented in Fig.~\ref{fig:ExD_neuro}. Scale bar, 100 \si{\micro\meter}.
 }%
\end{video}

Taken together, the above analyses demonstrate that ExD-SPIM offers optimal use of the photon budget, collecting more fluorescence by using a high-NA objective lens but extending the detection DOF to match the light-sheet thickness. The layer-cake phase mask used here does introduce a slight drop in peak intensity, but this loss is more than offset by the increased integrated signal collected axially. The improvement in signal collection increases with increasing NA (roughly proportional to NA$^2$), but the use of the phase mask allows us to harvest this increased efficiency without suffering from the narrow DOF that typically accompanies increasing NA \cite{Hect}.

ExD-SPIM offers an optimal compromise for light-sheet imaging of large specimens, permitting us to escape from the reduced performance of moderate-to-low NA optics of SPIM setups designed for imaging large samples \cite{Tomer_sped,Voleti_2019,Vettenburg_2015,Hosny_2020}. The results presented here show the improved photon efficiency from extended DOF at high NA, especially in low-light regimes where every photon counts. This allows ExD-SPIM to acquire images with significantly less laser excitation power, generating the same SNR with less photodamage, which is critical for studying light-sensitive systems or cellular dynamics over long timescales \cite{Lee_2017}. The simple modification needed to accomplish ExD-SPIM using a pupil phase mask avoids the expense, extensive optical redesign, or the wavelength-dependent phase modulation of other approaches, making it easily adaptable to multicolor fluorescence microscopy and scalable for high-throughput imaging.\\ \newline

We thank the David Prober Lab (California Institute of Technology) for early discussions that motivated the need to increase the photon detection efficiency of SPIM. We also thank Richard Kershaw (Hamamatsu Photonics) for allowing us to test the layer-cake phase mask. Our work was supported in part by the US National Science Foundation (1608744, 1650406, 1828793, PHY-1734030), the National Institutes of Health (1R01MH107238-01), the Human Frontier Science Program (RGP0008/2017), and the Jet Propulsion Laboratory (1632330); K. Keomanee-Dizon was supported in part by the Alfred E. Mann Doctoral Fellowship, and the Robert H. Dicke Fellowship in Experimental Physics from Princeton University.\newline

K.K.D., S.E.F., and T.V.T. have pending intellectual property related to the technology described here (patent application PCT/US2021/022198).

\section{Methods}

\subsection{Theory and numerical simulations}
\label{sec:ExD_sim_methods}
In its current configuration, ExD-SPIM implements a pupil-plane layer-cake phase mask \cite{Abrahamsson_2006} to match the DOF of a high-NA detection objective to the light-sheet thickness used. The layer-cake phase mask consists of multiple concentric glass discs (shaped like a wedding cake), each of which introduces an optical path length difference $\delta$ between adjacent zones:
$\delta=t(n-1)$,
where $t$ is the thickness of the layer and $n$ is the refractive index of the material. When $\delta$ is much longer than the short ($\sim\!\!1$ \si{\micro\meter}) coherence length of the fluorescence light (e.g., $\geq$ 50 \si{\micro\meter}), interference between each of the zones is eliminated. Each of the zones are thereby mutually incoherent, acting as a series of independent sub-apertures that form different foci with the same (elongated) axial extent, which add together incoherently to produce an effective focus, while largely maintaining its intrinsic lateral extent (Fig.~\ref{fig:layercake}).

We simulate the electromagnetic focus field of an objective using the vectorial Debye diffraction integral with the Fourier transform method
\cite{Leutenegger_2006}
given in Chen \textit{et al}.
\cite{ChenChakraborty_2020}. We do this by considering a continuum of wavevectors passed through an objective. These wavevectors lie on a (Ewald) spherical cap and have a magnitude $k=n\lambda$, where  $\lambda$ is wavelength of light; lateral ($x$-$y$) and axial ($x$-$z$) spatial features map to the set of wavevectors along $k\textsubscript{x}$ and $k\textsubscript{z}$, respectively. The set of $k\textsubscript{x}$ wavevectors is given by 2$k\sin(\alpha \textsubscript{2})$; the set of $k\textsubscript{z}$ wavevectors is defined as $k(\cos(\alpha \textsubscript{1}) - \cos(\alpha \textsubscript{2}))$. For a full spherical cap, $\alpha \textsubscript{1} = 0$ and hence
\begin{equation}
    \label{eq:dz}
    d\textsubscript{z} = \frac{\lambda}{n(1-\cos(\alpha \textsubscript{2}))}.
\end{equation}
We can now begin to geometrically map sub-apertures of the phase mask to wavevector sets. The central layer consists of a typical Gaussian-like focus so that the axial extent follows Eq.~\ref{eq:dz}. The other sub-apertures produce Bessel-like foci, where $\alpha \textsubscript{1} \neq 0$, so we must write
\begin{equation}
    \label{eq:dz_Bessel}
    d\textsubscript{z} = \frac{\lambda}{n(\cos(\alpha \textsubscript{1})-\cos(\alpha \textsubscript{2}))}.
\end{equation}
Since $\alpha = \arcsin(\frac{\mathit{NA}}{n})$, we can rewrite Eq.~\ref{eq:dz_Bessel} in terms of NA:
\begin{widetext}
\begin{equation}
    \label{eq:dz_Bessel_NA}
    d\textsubscript{z} = \frac{\lambda}{n \left[\cos \left(\arcsin \left(\frac{\mathit{NA} \textsubscript{1}}{n} \right) \right)-\cos \left(\arcsin \left(\frac{\mathit{NA} \textsubscript{2}}{n} \right) \right) \right]}.
\end{equation}
\end{widetext}

Given the back pupil of the objective is 2$\cdot f \cdot \mathit{NA}$, where $f$ is the back focal length of the objective, we can convert NAs to physical mask dimensions, or physical mask dimensions to NAs when the magnification between them is properly accounted for. From the above expressions, we can determine the required aperture diameter for a given DOF and number of annular sub-apertures. We start by calculating the inner most sub-aperture diameter (or equivalently, NA) that will produce a Gaussian-like focus with the desired DOF. Then, we calculate the first annular sub-aperture to have an inner diameter equivalent to the diameter of the central, low-NA, Gaussian focus, with an outer-ring diameter such that the Bessel-like focus has the same DOF using Eq.~\ref{eq:dz_Bessel_NA}. This is repeated so that the previous outer diameter is equivalent to the next inner diameter of the annular sub-apertures until all annuli are considered or the maximum available NA is filled. Each annular sub-aperture produces independent 3D electric fields; the modulus of these fields are squared to produce intensity PSFs. As discussed above, each PSF is mutually incoherent to the others and can thus be summed to form the effective detection PSF (Fig.~\ref{fig:layercake}). Theoretical simulations, where possible, match experimental parameters.

ExD-SPIM detection PSF: $\lambda\textsubscript{em}=525$ nm, $\mathit{NA}\textsubscript{det}=0.8$, $n=1.33$, and $L=4$, where $L$ is the number of annular sub-apertures in the layer-cake phase mask. This yielded a DOF of 8 \si{\micro\meter}.

High-NA detection PSF: $\lambda\textsubscript{em}=525$ nm, $\mathit{NA}\textsubscript{det}=0.8$, and $n=1.33$; this yielded a DOF of 2 \si{\micro\meter}.

Low-NA detection PSF: $\lambda\textsubscript{em}=525$ nm, $\mathit{NA}\textsubscript{det}=0.41$, and $n=1.33$; this yielded a DOF of 8 \si{\micro\meter}.

Excitation light-sheet: $\lambda\textsubscript{em}=525$ nm, $\mathit{NA}\textsubscript{det}=0.02$, $n=1.33$; this yielded a 6-\si{\micro\meter}-thick Gaussian light-sheet.

Overall PSFs were calculated by multiplying the detection PSF with the light-sheet PSF.

\subsection{Microscope optics}
\label{sec:ExD_optics}
The optical microscope was based on an existing light-sheet setup \cite{Madaan_2021}, with modifications to provide extension of the detection depth of field (Fig.~\ref{fig:ExD_optics}). The light-sheet module is described in detail in \cite{Madaan_2021}. We replaced the illumination lens with a 5$\times$, 0.1 NA, long-working distance objective lens (LMPLN5XIR, Olympus).

The detection module consisted of a 40$\times$, 0.8 NA, water-immersion objective lens (N40X-NIR, Nikon), mounted on a piezoelectric collar (P-725.4CD, Physik Instrumente); a set of tube lenses (200 mm focal length, TTL200, Thorlabs; 200 mm focal length, 49-286, Edmund Optics) that relayed the back pupil of the detection objective to the layer-cake phase mask (10 mm diameter aperture, 5 layers, A12802-35-040, Hamamatsu Photonics) used to extend the DOF. We under-filled the phase mask and used only 4 of the innermost layers, extending the DOF accordingly (Methods Section~\ref{sec:ExD_sim_methods}). A bandpass filter (525 $\pm$ 50 nm) follows the phase mask to separate the excitation light, and a tube lens (150 mm focal length, AC-508-150-A, Thorlabs) that forms the primary image at an sCMOS camera (Zyla 5.5, Andor) with an effective magnification of 30$\times$. For neural activity imaging, the fluorescence was imaged onto the camera at lower magnification (effectively 9.45$\times$) with a different tube lens (75 mm focal length, AC508-075-A-ML, Thorlabs).

The high NA mode was achieved simply by removing the phase mask. To switch to the low NA mode, the aperture of an iris positioned at the conjugate pupil plane was closed down to achieve an NA of $\sim0.41$.

\subsection{Sample preparation and imaging conditions}
\label{sec:ExD_sample}
Sub-diffractive fluorescent beads (175 $\pm$ 5 nm diameter; PS-Speck Microscope Point Source Kit, P7220, Molecular Probes) were diluted (1:1000) and embedded in a mixture of 2\% agarose, using the caddy-dive bar system described in Ref.~\cite{Keomanee-Dizon_2020}. The beads were then submerged in an imaging chamber filled with 60 mL of distilled water. A concentration of 10\% Iodixanol was added to the imaging chamber and left to diffuse throughout the agarose-embedded beads and imaging chamber overnight, prior to imaging. Iodixanol helped alleviate the index of refractive mismatch between the agarose and water \cite{Boothe_2017}.\\

Zebrafish were raised and maintained as described in
Ref. \cite{Westerfield},
in strict accordance with the recommendations in the Guide for the Care and Use of Laboratory Animals by University of Southern California, where the protocol was approved by the Institutional Animal Care and Use Committee (IACUC). All zebrafish lines used are available from ZIRC (\url{http://zebrafish.org}). Zebrafish embryos were collected from mating of appropriate adult fish (AB/TL strain) and raised in egg water (60 $\mu$g L$^{-3}$ of stock salts in distilled water) at 28.5$^{\circ}$C. At 20 hpf, 1-phenyl-2-thiourea (PTU) (30 mg L$^{-1}$) was added to the egg water to reduce pigmentation in the animals.

The imaging chamber was filled with 30\% Danieau solution (1740 mM NaCl, 21 mM KCl, 12 mM MgSO\textsubscript{4}$\cdot$7H\textsubscript{2}O, 18 mM Ca(NO\textsubscript{3})\textsubscript{2}, 150 mM HEPES). Fish were embedded in 2\% agarose as described in Ref.~\cite{Keomanee-Dizon_2020}. To measure photobleaching rates, cranial vasculature 4-5-dpf \textit{Tg(kdrl:eGFP)} larvae were used. The laser power was adjusted to reach similar signal levels (maximum and mean intensities) for each modality. Generally, this required 3- to 5-fold more laser power at low NA to reach the same SNR as the high NA systems, depending on the fish. Typical laser powers used for ExD-SPIM and high-NA SPIM ranged from 160 to 180 \si{\micro\watt} (measured at the back of the illumination objective); low-NA SPIM ranged from 0.45 to 0.75 mW. Exposure time was fixed at 200 ms for these experiments. We searched for similar features within the sample and acquired $N=2$ datasets for each modality.

For whole-brain neural activity imaging, 5-dpf zebrafish larvae expressing the nuclear-localized pan-neuro fluorescent calcium indicators, \textit{Tg(elavl3:H2B-GCaMP6s)}, were used. Imaging was performed over a $400\times800$ \si{\micro\meter}$^{2}$ region of the same specimen with 3.5 ms exposure ($\sim\!\!141$ frames per second) and $\sim\!\!150$ \si{\micro\watt} of laser power for all modalities. To cover a 200-\si{\micro\meter} $z$-depth of nearly the entire zebrafish brain, the $z$-scanning galvo was synchronized with the objective piezoelectric collar \cite{Keomanee-Dizon_2020}, and 25 image slices spaced 8 \si{\micro\meter} apart were recorded. We did not observe any appreciable signs of fluorescence photobleaching or phototoxicity over 10 minutes of recording from any of the modalities.

\subsection{Quantifying integrated signal-to-noise ratio}
\label{sec:SNR}
To quantify the relative integrated SNR in Fig.~\ref{fig:ExD_beads}d, we cropped the same ($N=6$) beads from $x$-$z$ summed-intensity projections for each mode. We define the relative integrated SNR as
\begin{equation}
    \label{eq:snr}
    \textrm{Relative integrated SNR} = \frac{S-\bar{b}}{\sigma\textsubscript{b}},
\end{equation}
where $S$ is the total integrated signal from an ovoid fit to the bead (size of ovoid is equivalent to the mean FWHM of the PSF for the respective mode), $\bar{b}$ is the total integrated background signal (measured in area devoid of bead, with the same ovoid used to calculate $S$), and $\sigma\textsubscript{b}$ is the standard deviation of the background.


\bibliography{references}

\begin{thebibliography}{33}%
\makeatletter
\providecommand \@ifxundefined [1]{%
 \@ifx{#1\undefined}
}%
\providecommand \@ifnum [1]{%
 \ifnum #1\expandafter \@firstoftwo
 \else \expandafter \@secondoftwo
 \fi
}%
\providecommand \@ifx [1]{%
 \ifx #1\expandafter \@firstoftwo
 \else \expandafter \@secondoftwo
 \fi
}%
\providecommand \natexlab [1]{#1}%
\providecommand \enquote  [1]{``#1''}%
\providecommand \bibnamefont  [1]{#1}%
\providecommand \bibfnamefont [1]{#1}%
\providecommand \citenamefont [1]{#1}%
\providecommand \href@noop [0]{\@secondoftwo}%
\providecommand \href [0]{\begingroup \@sanitize@url \@href}%
\providecommand \@href[1]{\@@startlink{#1}\@@href}%
\providecommand \@@href[1]{\endgroup#1\@@endlink}%
\providecommand \@sanitize@url [0]{\catcode `\\12\catcode `\$12\catcode
  `\&12\catcode `\#12\catcode `\^12\catcode `\_12\catcode `\%12\relax}%
\providecommand \@@startlink[1]{}%
\providecommand \@@endlink[0]{}%
\providecommand \url  [0]{\begingroup\@sanitize@url \@url }%
\providecommand \@url [1]{\endgroup\@href {#1}{\urlprefix }}%
\providecommand \urlprefix  [0]{URL }%
\providecommand \Eprint [0]{\href }%
\providecommand \doibase [0]{https://doi.org/}%
\providecommand \selectlanguage [0]{\@gobble}%
\providecommand \bibinfo  [0]{\@secondoftwo}%
\providecommand \bibfield  [0]{\@secondoftwo}%
\providecommand \translation [1]{[#1]}%
\providecommand \BibitemOpen [0]{}%
\providecommand \bibitemStop [0]{}%
\providecommand \bibitemNoStop [0]{.\EOS\space}%
\providecommand \EOS [0]{\spacefactor3000\relax}%
\providecommand \BibitemShut  [1]{\csname bibitem#1\endcsname}%
\let\auto@bib@innerbib\@empty
\bibitem [{\citenamefont {Power}\ and\ \citenamefont
  {Huisken}(2017)}]{Power_Huisken}%
  \BibitemOpen
  \bibfield  {author} {\bibinfo {author} {\bibfnamefont {R.~M.}\ \bibnamefont
  {Power}}\ and\ \bibinfo {author} {\bibfnamefont {J.}~\bibnamefont
  {Huisken}},\ }\bibfield  {title} {\bibinfo {title} {A guide to light-sheet
  fluorescence microscopy for multiscale imaging},\ }\href@noop {} {\bibfield
  {journal} {\bibinfo  {journal} {Nature Methods}\ }\textbf {\bibinfo {volume}
  {14}},\ \bibinfo {pages} {360} (\bibinfo {year} {2017})}\BibitemShut
  {NoStop}%
\bibitem [{\citenamefont {Siedentopf}\ and\ \citenamefont
  {Zsigmondy}(1902)}]{Siedentopf_Zsigmondy}%
  \BibitemOpen
  \bibfield  {author} {\bibinfo {author} {\bibfnamefont {H.}~\bibnamefont
  {Siedentopf}}\ and\ \bibinfo {author} {\bibfnamefont {R.}~\bibnamefont
  {Zsigmondy}},\ }\bibfield  {title} {\bibinfo {title} {On visualization and
  sizing of ultramicoscopic particles, with particular application to gold ruby
  glasses},\ }\href@noop {} {\bibfield  {journal} {\bibinfo  {journal} {Annalen
  der Physik}\ }\textbf {\bibinfo {volume} {315}},\ \bibinfo {pages} {1}
  (\bibinfo {year} {1902})}\BibitemShut {NoStop}%
\bibitem [{\citenamefont {Huisken}\ \emph {et~al.}(2004)\citenamefont
  {Huisken}, \citenamefont {Swoger}, \citenamefont {Bene}, \citenamefont
  {Wittbrodt},\ and\ \citenamefont {Stelzer}}]{Huisken_2004}%
  \BibitemOpen
  \bibfield  {author} {\bibinfo {author} {\bibfnamefont {J.}~\bibnamefont
  {Huisken}}, \bibinfo {author} {\bibfnamefont {J.}~\bibnamefont {Swoger}},
  \bibinfo {author} {\bibfnamefont {F.~D.}\ \bibnamefont {Bene}}, \bibinfo
  {author} {\bibfnamefont {J.}~\bibnamefont {Wittbrodt}},\ and\ \bibinfo
  {author} {\bibfnamefont {E.}~\bibnamefont {Stelzer}},\ }\bibfield  {title}
  {\bibinfo {title} {Optical sectioning deep inside live embryos by selective
  plane illumination microscopy},\ }\href@noop {} {\bibfield  {journal}
  {\bibinfo  {journal} {Science}\ }\textbf {\bibinfo {volume} {305}},\ \bibinfo
  {pages} {1007} (\bibinfo {year} {2004})}\BibitemShut {NoStop}%
\bibitem [{\citenamefont {Pawley}(2006)}]{Pawley}%
  \BibitemOpen
  \bibfield  {author} {\bibinfo {author} {\bibfnamefont {J.~B.}\ \bibnamefont
  {Pawley}},\ }\href@noop {} {\emph {\bibinfo {title} {Handbook of Confocal
  Microscopy}}},\ \bibinfo {edition} {3rd}\ ed.\ (\bibinfo  {publisher}
  {Springer Science and Business Media},\ \bibinfo {year} {2006})\BibitemShut
  {NoStop}%
\bibitem [{\citenamefont {Supatto}\ \emph {et~al.}(2011)\citenamefont
  {Supatto}, \citenamefont {Truong}, \citenamefont {D{\'e}barre},\ and\
  \citenamefont {Beaurepaire}}]{Supatto_2011}%
  \BibitemOpen
  \bibfield  {author} {\bibinfo {author} {\bibfnamefont {W.}~\bibnamefont
  {Supatto}}, \bibinfo {author} {\bibfnamefont {T.~V.}\ \bibnamefont {Truong}},
  \bibinfo {author} {\bibfnamefont {D.}~\bibnamefont {D{\'e}barre}},\ and\
  \bibinfo {author} {\bibfnamefont {E.}~\bibnamefont {Beaurepaire}},\
  }\bibfield  {title} {\bibinfo {title} {Advances in multiphoton microscopy for
  imaging embryos},\ }\href@noop {} {\bibfield  {journal} {\bibinfo  {journal}
  {Current Opinion in Genetics \& Development}\ }\textbf {\bibinfo {volume}
  {21}},\ \bibinfo {pages} {538} (\bibinfo {year} {2011})}\BibitemShut
  {NoStop}%
\bibitem [{\citenamefont {Garc{\'e}s-Ch{\'a}vez}\ \emph
  {et~al.}(2002)\citenamefont {Garc{\'e}s-Ch{\'a}vez}, \citenamefont {McGloin},
  \citenamefont {Melville}, \citenamefont {Sibbett},\ and\ \citenamefont
  {Dholakia}}]{Garces-Chavez_2002}%
  \BibitemOpen
  \bibfield  {author} {\bibinfo {author} {\bibfnamefont {V.}~\bibnamefont
  {Garc{\'e}s-Ch{\'a}vez}}, \bibinfo {author} {\bibfnamefont {D.}~\bibnamefont
  {McGloin}}, \bibinfo {author} {\bibfnamefont {H.}~\bibnamefont {Melville}},
  \bibinfo {author} {\bibfnamefont {W.}~\bibnamefont {Sibbett}},\ and\ \bibinfo
  {author} {\bibfnamefont {K.}~\bibnamefont {Dholakia}},\ }\bibfield  {title}
  {\bibinfo {title} {Simultaneous micromanipulation in multiple planes using a
  self-reconstructing light beam},\ }\href@noop {} {\bibfield  {journal}
  {\bibinfo  {journal} {Nature}\ }\textbf {\bibinfo {volume} {419}} (\bibinfo
  {year} {2002})}\BibitemShut {NoStop}%
\bibitem [{\citenamefont {Fahrbach}\ \emph {et~al.}(2010)\citenamefont
  {Fahrbach}, \citenamefont {Simon},\ and\ \citenamefont
  {Rohrbach}}]{Fahrbach_2010}%
  \BibitemOpen
  \bibfield  {author} {\bibinfo {author} {\bibfnamefont {F.~O.}\ \bibnamefont
  {Fahrbach}}, \bibinfo {author} {\bibfnamefont {P.}~\bibnamefont {Simon}},\
  and\ \bibinfo {author} {\bibfnamefont {A.}~\bibnamefont {Rohrbach}},\
  }\bibfield  {title} {\bibinfo {title} {Microscopy with selfreconstructing
  beams},\ }\href@noop {} {\bibfield  {journal} {\bibinfo  {journal} {Nature
  Photonics}\ }\textbf {\bibinfo {volume} {4}},\ \bibinfo {pages} {780}
  (\bibinfo {year} {2010})}\BibitemShut {NoStop}%
\bibitem [{\citenamefont {Planchon}\ \emph {et~al.}(2011)\citenamefont
  {Planchon}, \citenamefont {Gao}, \citenamefont {Milkie}, \citenamefont
  {Davidson}, \citenamefont {Galbraith}, \citenamefont {Galbraith},\ and\
  \citenamefont {Betzig}}]{Planchon_2011}%
  \BibitemOpen
  \bibfield  {author} {\bibinfo {author} {\bibfnamefont {T.}~\bibnamefont
  {Planchon}}, \bibinfo {author} {\bibfnamefont {L.}~\bibnamefont {Gao}},
  \bibinfo {author} {\bibfnamefont {D.~E.}\ \bibnamefont {Milkie}}, \bibinfo
  {author} {\bibfnamefont {M.~W.}\ \bibnamefont {Davidson}}, \bibinfo {author}
  {\bibfnamefont {J.~A.}\ \bibnamefont {Galbraith}}, \bibinfo {author}
  {\bibfnamefont {C.~G.}\ \bibnamefont {Galbraith}},\ and\ \bibinfo {author}
  {\bibfnamefont {E.}~\bibnamefont {Betzig}},\ }\bibfield  {title} {\bibinfo
  {title} {Rapid three-dimensional isotropic imaging of living cells using
  bessel beam plane illumination},\ }\href@noop {} {\bibfield  {journal}
  {\bibinfo  {journal} {Nature Methods}\ }\textbf {\bibinfo {volume} {8}},\
  \bibinfo {pages} {417} (\bibinfo {year} {2011})}\BibitemShut {NoStop}%
\bibitem [{\citenamefont {Vettenburg}\ \emph {et~al.}(2014)\citenamefont
  {Vettenburg}, \citenamefont {Dalgarno}, \citenamefont {Nylk}, \citenamefont
  {Coll-Llad{\'o}}, \citenamefont {Ferrier}, \citenamefont {{\v C}i{\v
  z}m{\'a}r}, \citenamefont {Gunn-Moore},\ and\ \citenamefont
  {Dholakia}}]{Vettenburg_2015}%
  \BibitemOpen
  \bibfield  {author} {\bibinfo {author} {\bibfnamefont {T.}~\bibnamefont
  {Vettenburg}}, \bibinfo {author} {\bibfnamefont {H.~I.~C.}\ \bibnamefont
  {Dalgarno}}, \bibinfo {author} {\bibfnamefont {J.}~\bibnamefont {Nylk}},
  \bibinfo {author} {\bibfnamefont {C.}~\bibnamefont {Coll-Llad{\'o}}},
  \bibinfo {author} {\bibfnamefont {D.~E.~K.}\ \bibnamefont {Ferrier}},
  \bibinfo {author} {\bibfnamefont {T.}~\bibnamefont {{\v C}i{\v z}m{\'a}r}},
  \bibinfo {author} {\bibfnamefont {F.~J.}\ \bibnamefont {Gunn-Moore}},\ and\
  \bibinfo {author} {\bibfnamefont {K.}~\bibnamefont {Dholakia}},\ }\bibfield
  {title} {\bibinfo {title} {Light-sheet microscopy using an airy beam},\
  }\href@noop {} {\bibfield  {journal} {\bibinfo  {journal} {Nature Methods}\
  }\textbf {\bibinfo {volume} {11}},\ \bibinfo {pages} {541} (\bibinfo {year}
  {2014})}\BibitemShut {NoStop}%
\bibitem [{\citenamefont {Hosny}\ \emph {et~al.}(2020)\citenamefont {Hosny},
  \citenamefont {Seyforth}, \citenamefont {Spickermann}, \citenamefont
  {Mitchell}, \citenamefont {Almada}, \citenamefont {Chesters}, \citenamefont
  {Mitchell}, \citenamefont {Chennell}, \citenamefont {Vernon}, \citenamefont
  {Cho}, \citenamefont {Srivastava}, \citenamefont {Forster},\ and\
  \citenamefont {Vettenburg}}]{Hosny_2020}%
  \BibitemOpen
  \bibfield  {author} {\bibinfo {author} {\bibfnamefont {N.~A.}\ \bibnamefont
  {Hosny}}, \bibinfo {author} {\bibfnamefont {J.~A.}\ \bibnamefont {Seyforth}},
  \bibinfo {author} {\bibfnamefont {G.}~\bibnamefont {Spickermann}}, \bibinfo
  {author} {\bibfnamefont {T.~J.}\ \bibnamefont {Mitchell}}, \bibinfo {author}
  {\bibfnamefont {P.}~\bibnamefont {Almada}}, \bibinfo {author} {\bibfnamefont
  {R.}~\bibnamefont {Chesters}}, \bibinfo {author} {\bibfnamefont {S.~J.}\
  \bibnamefont {Mitchell}}, \bibinfo {author} {\bibfnamefont {G.}~\bibnamefont
  {Chennell}}, \bibinfo {author} {\bibfnamefont {A.~C.}\ \bibnamefont
  {Vernon}}, \bibinfo {author} {\bibfnamefont {K.}~\bibnamefont {Cho}},
  \bibinfo {author} {\bibfnamefont {D.~P.}\ \bibnamefont {Srivastava}},
  \bibinfo {author} {\bibfnamefont {R.}~\bibnamefont {Forster}},\ and\ \bibinfo
  {author} {\bibfnamefont {T.}~\bibnamefont {Vettenburg}},\ }\bibfield  {title}
  {\bibinfo {title} {Planar airy beam light-sheet for two-photon microscopy},\
  }\href@noop {} {\bibfield  {journal} {\bibinfo  {journal} {Biomedical Optics
  Express}\ }\textbf {\bibinfo {volume} {11}},\ \bibinfo {pages} {3927}
  (\bibinfo {year} {2020})}\BibitemShut {NoStop}%
\bibitem [{\citenamefont {Chen}\ \emph {et~al.}(2014)\citenamefont {Chen},
  \citenamefont {Legant}, \citenamefont {Wang}, \citenamefont {Shao},
  \citenamefont {Milkie}, \citenamefont {Daniel}, \citenamefont {Davidson},
  \citenamefont {Michael}, \citenamefont {Janetopoulos}, \citenamefont {Wu},
  \citenamefont {Hammer}, \citenamefont {Liu}, \citenamefont {English},
  \citenamefont {Mimori-Kiyosue}, \citenamefont {Romero}, \citenamefont
  {Ritter}, \citenamefont {Lippincott-Schwartz}, \citenamefont {FritzLaylin},
  \citenamefont {Mullins}, \citenamefont {Mitchell}, \citenamefont {Bembenek},
  \citenamefont {Reymann}, \citenamefont {B{\"o}hme}, \citenamefont {Grill},
  \citenamefont {Wang}, \citenamefont {Seydoux}, \citenamefont {Tulu},
  \citenamefont {Kiehart},\ and\ \citenamefont {Betzig}}]{Chen_2014}%
  \BibitemOpen
  \bibfield  {author} {\bibinfo {author} {\bibfnamefont {B.-C.}\ \bibnamefont
  {Chen}}, \bibinfo {author} {\bibfnamefont {W.~R.}\ \bibnamefont {Legant}},
  \bibinfo {author} {\bibfnamefont {K.}~\bibnamefont {Wang}}, \bibinfo {author}
  {\bibfnamefont {L.}~\bibnamefont {Shao}}, \bibinfo {author} {\bibfnamefont
  {D.~E.}\ \bibnamefont {Milkie}}, \bibinfo {author} {\bibfnamefont
  {E.}~\bibnamefont {Daniel}}, \bibinfo {author} {\bibfnamefont {M.~W.}\
  \bibnamefont {Davidson}}, \bibinfo {author} {\bibfnamefont {W.}~\bibnamefont
  {Michael}}, \bibinfo {author} {\bibfnamefont {C.}~\bibnamefont
  {Janetopoulos}}, \bibinfo {author} {\bibfnamefont {X.~S.}\ \bibnamefont
  {Wu}}, \bibinfo {author} {\bibfnamefont {J.~A.}\ \bibnamefont {Hammer}},
  \bibinfo {author} {\bibfnamefont {Z.}~\bibnamefont {Liu}}, \bibinfo {author}
  {\bibfnamefont {B.~P.}\ \bibnamefont {English}}, \bibinfo {author}
  {\bibfnamefont {Y.}~\bibnamefont {Mimori-Kiyosue}}, \bibinfo {author}
  {\bibfnamefont {D.~P.}\ \bibnamefont {Romero}}, \bibinfo {author}
  {\bibfnamefont {A.~T.}\ \bibnamefont {Ritter}}, \bibinfo {author}
  {\bibfnamefont {J.}~\bibnamefont {Lippincott-Schwartz}}, \bibinfo {author}
  {\bibfnamefont {L.}~\bibnamefont {FritzLaylin}}, \bibinfo {author}
  {\bibfnamefont {R.~D.}\ \bibnamefont {Mullins}}, \bibinfo {author}
  {\bibfnamefont {D.~M.}\ \bibnamefont {Mitchell}}, \bibinfo {author}
  {\bibfnamefont {J.~N.}\ \bibnamefont {Bembenek}}, \bibinfo {author}
  {\bibfnamefont {A.~C.}\ \bibnamefont {Reymann}}, \bibinfo {author}
  {\bibfnamefont {R.}~\bibnamefont {B{\"o}hme}}, \bibinfo {author}
  {\bibfnamefont {S.~W.}\ \bibnamefont {Grill}}, \bibinfo {author}
  {\bibfnamefont {J.~T.}\ \bibnamefont {Wang}}, \bibinfo {author}
  {\bibfnamefont {G.}~\bibnamefont {Seydoux}}, \bibinfo {author} {\bibfnamefont
  {U.~S.}\ \bibnamefont {Tulu}}, \bibinfo {author} {\bibfnamefont {D.~P.}\
  \bibnamefont {Kiehart}},\ and\ \bibinfo {author} {\bibfnamefont
  {E.}~\bibnamefont {Betzig}},\ }\bibfield  {title} {\bibinfo {title} {Lattice
  light-sheet microscopy: Imaging molecules to embryos at high spatiotemporal
  resolution},\ }\href@noop {} {\bibfield  {journal} {\bibinfo  {journal}
  {Science}\ }\textbf {\bibinfo {volume} {346}},\ \bibinfo {pages} {1257998}
  (\bibinfo {year} {2014})}\BibitemShut {NoStop}%
\bibitem [{\citenamefont {Gao}\ \emph {et~al.}(2019)\citenamefont {Gao},
  \citenamefont {Asano}, \citenamefont {Upadhyayula}, \citenamefont {Pisarev},
  \citenamefont {Milkie}, \citenamefont {Liu}, \citenamefont {Singh},
  \citenamefont {Graves}, \citenamefont {Huynh}, \citenamefont {Zhao},
  \citenamefont {Bogovic}, \citenamefont {Colonell}, \citenamefont {Ott},
  \citenamefont {Zugates}, \citenamefont {Tappan}, \citenamefont {Rodriguez},
  \citenamefont {Mosaliganti}, \citenamefont {Sheu}, \citenamefont {Pasolli},
  \citenamefont {Pang}, \citenamefont {Xu}, \citenamefont {Megason},
  \citenamefont {Hess}, \citenamefont {Lippincott-Schwartz}, \citenamefont
  {Hantman}, \citenamefont {Rubin}, \citenamefont {Kirchhausen}, \citenamefont
  {Saalfeld}, \citenamefont {Aso}, \citenamefont {Boyden},\ and\ \citenamefont
  {Betzig}}]{Gao_2019}%
  \BibitemOpen
  \bibfield  {author} {\bibinfo {author} {\bibfnamefont {R.}~\bibnamefont
  {Gao}}, \bibinfo {author} {\bibfnamefont {S.~M.}\ \bibnamefont {Asano}},
  \bibinfo {author} {\bibfnamefont {U.}~\bibnamefont {Upadhyayula}}, \bibinfo
  {author} {\bibfnamefont {I.}~\bibnamefont {Pisarev}}, \bibinfo {author}
  {\bibfnamefont {D.~E.}\ \bibnamefont {Milkie}}, \bibinfo {author}
  {\bibfnamefont {T.-L.}\ \bibnamefont {Liu}}, \bibinfo {author} {\bibfnamefont
  {V.}~\bibnamefont {Singh}}, \bibinfo {author} {\bibfnamefont
  {A.}~\bibnamefont {Graves}}, \bibinfo {author} {\bibfnamefont {G.~H.}\
  \bibnamefont {Huynh}}, \bibinfo {author} {\bibfnamefont {Y.}~\bibnamefont
  {Zhao}}, \bibinfo {author} {\bibfnamefont {J.}~\bibnamefont {Bogovic}},
  \bibinfo {author} {\bibfnamefont {J.}~\bibnamefont {Colonell}}, \bibinfo
  {author} {\bibfnamefont {C.~M.}\ \bibnamefont {Ott}}, \bibinfo {author}
  {\bibfnamefont {C.}~\bibnamefont {Zugates}}, \bibinfo {author} {\bibfnamefont
  {S.}~\bibnamefont {Tappan}}, \bibinfo {author} {\bibfnamefont
  {A.}~\bibnamefont {Rodriguez}}, \bibinfo {author} {\bibfnamefont {K.~R.}\
  \bibnamefont {Mosaliganti}}, \bibinfo {author} {\bibfnamefont {S.-H.}\
  \bibnamefont {Sheu}}, \bibinfo {author} {\bibfnamefont {H.~A.}\ \bibnamefont
  {Pasolli}}, \bibinfo {author} {\bibfnamefont {S.}~\bibnamefont {Pang}},
  \bibinfo {author} {\bibfnamefont {C.~S.}\ \bibnamefont {Xu}}, \bibinfo
  {author} {\bibfnamefont {S.~G.}\ \bibnamefont {Megason}}, \bibinfo {author}
  {\bibfnamefont {H.}~\bibnamefont {Hess}}, \bibinfo {author} {\bibfnamefont
  {J.}~\bibnamefont {Lippincott-Schwartz}}, \bibinfo {author} {\bibfnamefont
  {A.}~\bibnamefont {Hantman}}, \bibinfo {author} {\bibfnamefont {G.~M.}\
  \bibnamefont {Rubin}}, \bibinfo {author} {\bibfnamefont {T.}~\bibnamefont
  {Kirchhausen}}, \bibinfo {author} {\bibfnamefont {S.}~\bibnamefont
  {Saalfeld}}, \bibinfo {author} {\bibfnamefont {Y.}~\bibnamefont {Aso}},
  \bibinfo {author} {\bibfnamefont {E.~S.}\ \bibnamefont {Boyden}},\ and\
  \bibinfo {author} {\bibfnamefont {E.}~\bibnamefont {Betzig}},\ }\bibfield
  {title} {\bibinfo {title} {Cortical column and whole-brain imaging with
  molecular contrast and nanoscale resolution},\ }\href@noop {} {\ \textbf
  {\bibinfo {volume} {363}},\ \bibinfo {pages} {6424} (\bibinfo {year}
  {2019})}\BibitemShut {NoStop}%
\bibitem [{\citenamefont {Keller}\ \emph {et~al.}(2008)\citenamefont {Keller},
  \citenamefont {Schmidt}, \citenamefont {Wittbrodt},\ and\ \citenamefont
  {Stelzer}}]{Keller_2008}%
  \BibitemOpen
  \bibfield  {author} {\bibinfo {author} {\bibfnamefont {P.~J.}\ \bibnamefont
  {Keller}}, \bibinfo {author} {\bibfnamefont {A.~D.}\ \bibnamefont {Schmidt}},
  \bibinfo {author} {\bibfnamefont {J.}~\bibnamefont {Wittbrodt}},\ and\
  \bibinfo {author} {\bibfnamefont {E.~H.~K.}\ \bibnamefont {Stelzer}},\
  }\bibfield  {title} {\bibinfo {title} {Reconstruction of zebrafish early
  embryonic development by scanned light sheet microscopy},\ }\href@noop {}
  {\bibfield  {journal} {\bibinfo  {journal} {Science}\ }\textbf {\bibinfo
  {volume} {322}},\ \bibinfo {pages} {1065} (\bibinfo {year}
  {2008})}\BibitemShut {NoStop}%
\bibitem [{\citenamefont {Ahrens}\ \emph {et~al.}(2013)\citenamefont {Ahrens},
  \citenamefont {Orger}, \citenamefont {Robson}, \citenamefont {Li},\ and\
  \citenamefont {Keller}}]{Ahrens_2013}%
  \BibitemOpen
  \bibfield  {author} {\bibinfo {author} {\bibfnamefont {M.~B.}\ \bibnamefont
  {Ahrens}}, \bibinfo {author} {\bibfnamefont {M.~B.}\ \bibnamefont {Orger}},
  \bibinfo {author} {\bibfnamefont {D.~N.}\ \bibnamefont {Robson}}, \bibinfo
  {author} {\bibfnamefont {J.~M.}\ \bibnamefont {Li}},\ and\ \bibinfo {author}
  {\bibfnamefont {P.~J.}\ \bibnamefont {Keller}},\ }\bibfield  {title}
  {\bibinfo {title} {Whole-brain functional imaging at cellular resolution
  using light-sheet microscopy},\ }\href@noop {} {\bibfield  {journal}
  {\bibinfo  {journal} {Nature Methods}\ }\textbf {\bibinfo {volume} {10}},\
  \bibinfo {pages} {413} (\bibinfo {year} {2013})}\BibitemShut {NoStop}%
\bibitem [{\citenamefont {Wolf}\ \emph {et~al.}(2015)\citenamefont {Wolf},
  \citenamefont {Supatto}, \citenamefont {Debr{\'e}geas}, \citenamefont
  {Mahou}, \citenamefont {Kruglik}, \citenamefont {Sintes}, \citenamefont
  {Beaurepaire},\ and\ \citenamefont {Candelier}}]{Wolf_2015}%
  \BibitemOpen
  \bibfield  {author} {\bibinfo {author} {\bibfnamefont {S.}~\bibnamefont
  {Wolf}}, \bibinfo {author} {\bibfnamefont {W.}~\bibnamefont {Supatto}},
  \bibinfo {author} {\bibfnamefont {G.}~\bibnamefont {Debr{\'e}geas}}, \bibinfo
  {author} {\bibfnamefont {P.}~\bibnamefont {Mahou}}, \bibinfo {author}
  {\bibfnamefont {S.}~\bibnamefont {Kruglik}}, \bibinfo {author} {\bibfnamefont
  {J.~M.}\ \bibnamefont {Sintes}}, \bibinfo {author} {\bibfnamefont
  {E.}~\bibnamefont {Beaurepaire}},\ and\ \bibinfo {author} {\bibfnamefont
  {R.}~\bibnamefont {Candelier}},\ }\bibfield  {title} {\bibinfo {title}
  {Whole-brain functional imaging with two-photon light-sheet microscopy},\
  }\href@noop {} {\bibfield  {journal} {\bibinfo  {journal} {Nature Methods}\
  }\textbf {\bibinfo {volume} {12}},\ \bibinfo {pages} {379} (\bibinfo {year}
  {2015})}\BibitemShut {NoStop}%
\bibitem [{\citenamefont {Keomanee-Dizon}\ \emph {et~al.}(2020)\citenamefont
  {Keomanee-Dizon}, \citenamefont {Fraser},\ and\ \citenamefont
  {Truong}}]{Keomanee-Dizon_2020}%
  \BibitemOpen
  \bibfield  {author} {\bibinfo {author} {\bibfnamefont {K.}~\bibnamefont
  {Keomanee-Dizon}}, \bibinfo {author} {\bibfnamefont {S.~E.}\ \bibnamefont
  {Fraser}},\ and\ \bibinfo {author} {\bibfnamefont {T.~V.}\ \bibnamefont
  {Truong}},\ }\bibfield  {title} {\bibinfo {title} {A versatile, multi-laser
  twin-microscope system for light-sheet imaging},\ }\href@noop {} {\bibfield
  {journal} {\bibinfo  {journal} {Review of Scientific Instruments}\ }\textbf
  {\bibinfo {volume} {91}},\ \bibinfo {pages} {053703} (\bibinfo {year}
  {2020})}\BibitemShut {NoStop}%
\bibitem [{\citenamefont {Abrahamsson}\ \emph {et~al.}(2006)\citenamefont
  {Abrahamsson}, \citenamefont {Usawa},\ and\ \citenamefont
  {Gustafsson}}]{Abrahamsson_2006}%
  \BibitemOpen
  \bibfield  {author} {\bibinfo {author} {\bibfnamefont {S.}~\bibnamefont
  {Abrahamsson}}, \bibinfo {author} {\bibfnamefont {S.}~\bibnamefont {Usawa}},\
  and\ \bibinfo {author} {\bibfnamefont {M.}~\bibnamefont {Gustafsson}},\
  }\bibfield  {title} {\bibinfo {title} {A new approach to extended focus for
  high-speed high-resolution biological microscopy},\ }in\ \href@noop {} {\emph
  {\bibinfo {booktitle} {Three-Dimensional and Multidimensional Microscopy:
  Image Acquisition and Processing XIII}}}\ (\bibinfo {organization}
  {International Society for Optics and Photonics},\ \bibinfo {year} {2006})\
  p.\ \bibinfo {pages} {60900N}\BibitemShut {NoStop}%
\bibitem [{\citenamefont {Madaan}\ \emph {et~al.}(2021)\citenamefont {Madaan},
  \citenamefont {Keomanee-Dizon}, \citenamefont {Jones}, \citenamefont {Zhong},
  \citenamefont {Nadtochiy}, \citenamefont {Luu}, \citenamefont {Fraser},\ and\
  \citenamefont {Truong}}]{Madaan_2021}%
  \BibitemOpen
  \bibfield  {author} {\bibinfo {author} {\bibfnamefont {S.}~\bibnamefont
  {Madaan}}, \bibinfo {author} {\bibfnamefont {K.}~\bibnamefont
  {Keomanee-Dizon}}, \bibinfo {author} {\bibfnamefont {M.}~\bibnamefont
  {Jones}}, \bibinfo {author} {\bibfnamefont {C.}~\bibnamefont {Zhong}},
  \bibinfo {author} {\bibfnamefont {A.}~\bibnamefont {Nadtochiy}}, \bibinfo
  {author} {\bibfnamefont {P.}~\bibnamefont {Luu}}, \bibinfo {author}
  {\bibfnamefont {S.~E.}\ \bibnamefont {Fraser}},\ and\ \bibinfo {author}
  {\bibfnamefont {T.~V.}\ \bibnamefont {Truong}},\ }\bibfield  {title}
  {\bibinfo {title} {Single-objective selective-volume illumination microscopy
  enables high-contrast light-field imaging},\ }\href@noop {} {\bibfield
  {journal} {\bibinfo  {journal} {Optics Letters}\ }\textbf {\bibinfo {volume}
  {46}},\ \bibinfo {pages} {2860} (\bibinfo {year} {2021})}\BibitemShut
  {NoStop}%
\bibitem [{\citenamefont {Olarte}\ \emph {et~al.}(2015)\citenamefont {Olarte},
  \citenamefont {Artigas},\ and\ \citenamefont {Loza-Alvarez}}]{Olarte_2015}%
  \BibitemOpen
  \bibfield  {author} {\bibinfo {author} {\bibfnamefont {O.~E.}\ \bibnamefont
  {Olarte}}, \bibinfo {author} {\bibfnamefont {J.~A.~D.}\ \bibnamefont
  {Artigas}},\ and\ \bibinfo {author} {\bibfnamefont {P.}~\bibnamefont
  {Loza-Alvarez}},\ }\bibfield  {title} {\bibinfo {title} {Decoupled
  illumination detection in light-sheet microscopy for fast volumetric
  imaging},\ }\href@noop {} {\bibfield  {journal} {\bibinfo  {journal}
  {Optica}\ }\textbf {\bibinfo {volume} {2}},\ \bibinfo {pages} {702} (\bibinfo
  {year} {2015})}\BibitemShut {NoStop}%
\bibitem [{\citenamefont {Tomer}\ \emph {et~al.}(2015)\citenamefont {Tomer},
  \citenamefont {Lovett-Barron}, \citenamefont {Kauvar}, \citenamefont
  {Andalman}, \citenamefont {Burns}, \citenamefont {Sankaran}, \citenamefont
  {Grosenick}, \citenamefont {Broxton}, \citenamefont {Yang},\ and\
  \citenamefont {Deisseroth}}]{Tomer_sped}%
  \BibitemOpen
  \bibfield  {author} {\bibinfo {author} {\bibfnamefont {R.}~\bibnamefont
  {Tomer}}, \bibinfo {author} {\bibfnamefont {M.}~\bibnamefont
  {Lovett-Barron}}, \bibinfo {author} {\bibfnamefont {I.}~\bibnamefont
  {Kauvar}}, \bibinfo {author} {\bibfnamefont {A.}~\bibnamefont {Andalman}},
  \bibinfo {author} {\bibfnamefont {V.~M.}\ \bibnamefont {Burns}}, \bibinfo
  {author} {\bibfnamefont {S.}~\bibnamefont {Sankaran}}, \bibinfo {author}
  {\bibfnamefont {L.}~\bibnamefont {Grosenick}}, \bibinfo {author}
  {\bibfnamefont {M.}~\bibnamefont {Broxton}}, \bibinfo {author} {\bibfnamefont
  {S.}~\bibnamefont {Yang}},\ and\ \bibinfo {author} {\bibfnamefont
  {K.}~\bibnamefont {Deisseroth}},\ }\bibfield  {title} {\bibinfo {title} {Sped
  light sheet microscopy: Fast mapping of biological system structure and
  function},\ }\href@noop {} {\bibfield  {journal} {\bibinfo  {journal} {Cell}\
  }\textbf {\bibinfo {volume} {163}},\ \bibinfo {pages} {1796} (\bibinfo {year}
  {2015})}\BibitemShut {NoStop}%
\bibitem [{\citenamefont {Quirin}\ \emph {et~al.}(2016)\citenamefont {Quirin},
  \citenamefont {Vladimirov}, \citenamefont {Yang}, \citenamefont {Peterka},
  \citenamefont {Yuste},\ and\ \citenamefont {Ahrens}}]{Quirin_2016}%
  \BibitemOpen
  \bibfield  {author} {\bibinfo {author} {\bibfnamefont {S.}~\bibnamefont
  {Quirin}}, \bibinfo {author} {\bibfnamefont {N.}~\bibnamefont {Vladimirov}},
  \bibinfo {author} {\bibfnamefont {C.-T.}\ \bibnamefont {Yang}}, \bibinfo
  {author} {\bibfnamefont {D.~S.}\ \bibnamefont {Peterka}}, \bibinfo {author}
  {\bibfnamefont {R.}~\bibnamefont {Yuste}},\ and\ \bibinfo {author}
  {\bibfnamefont {M.~B.}\ \bibnamefont {Ahrens}},\ }\bibfield  {title}
  {\bibinfo {title} {Calcium imaging of neural circuits with extended
  depth-of-field light-sheet microscopy},\ }\href@noop {} {\bibfield  {journal}
  {\bibinfo  {journal} {Optics Letters}\ }\textbf {\bibinfo {volume} {41}},\
  \bibinfo {pages} {855} (\bibinfo {year} {2016})}\BibitemShut {NoStop}%
\bibitem [{\citenamefont {R.}\ and\ \citenamefont {Cathey}(1995)}]{Dowski}%
  \BibitemOpen
  \bibfield  {author} {\bibinfo {author} {\bibfnamefont {E.}~\bibnamefont
  {R.}}\ and\ \bibinfo {author} {\bibfnamefont {W.~T.}\ \bibnamefont
  {Cathey}},\ }\bibfield  {title} {\bibinfo {title} {Extended depth of field
  through wave-front coding},\ }\href@noop {} {\bibfield  {journal} {\bibinfo
  {journal} {Applied Optics}\ }\textbf {\bibinfo {volume} {34}},\ \bibinfo
  {pages} {1859} (\bibinfo {year} {1995})}\BibitemShut {NoStop}%
\bibitem [{\citenamefont {Fahrbach}\ \emph {et~al.}(2013)\citenamefont
  {Fahrbach}, \citenamefont {Voigt}, \citenamefont {Schmid}, \citenamefont
  {Helmchen},\ and\ \citenamefont {Huisken}}]{Fahrbach_2013}%
  \BibitemOpen
  \bibfield  {author} {\bibinfo {author} {\bibfnamefont {F.~O.}\ \bibnamefont
  {Fahrbach}}, \bibinfo {author} {\bibfnamefont {F.~F.}\ \bibnamefont {Voigt}},
  \bibinfo {author} {\bibfnamefont {B.}~\bibnamefont {Schmid}}, \bibinfo
  {author} {\bibfnamefont {F.}~\bibnamefont {Helmchen}},\ and\ \bibinfo
  {author} {\bibfnamefont {J.}~\bibnamefont {Huisken}},\ }\bibfield  {title}
  {\bibinfo {title} {Rapid 3d light-sheet microscopy with a tunable lens},\
  }\href@noop {} {\bibfield  {journal} {\bibinfo  {journal} {Optics Express}\
  }\textbf {\bibinfo {volume} {21}},\ \bibinfo {pages} {21010} (\bibinfo {year}
  {2013})}\BibitemShut {NoStop}%
\bibitem [{\citenamefont {Botcherby}\ \emph {et~al.}(2007)\citenamefont
  {Botcherby}, \citenamefont {Juskaitis}, \citenamefont {Booth},\ and\
  \citenamefont {Wilson}}]{Botcherby_2007}%
  \BibitemOpen
  \bibfield  {author} {\bibinfo {author} {\bibfnamefont {E.~J.}\ \bibnamefont
  {Botcherby}}, \bibinfo {author} {\bibfnamefont {R.}~\bibnamefont
  {Juskaitis}}, \bibinfo {author} {\bibfnamefont {M.~J.}\ \bibnamefont
  {Booth}},\ and\ \bibinfo {author} {\bibfnamefont {T.}~\bibnamefont
  {Wilson}},\ }\bibfield  {title} {\bibinfo {title} {Aberration-free optical
  refocusing in high numerical aperture microscopy},\ }\href@noop {} {\bibfield
   {journal} {\bibinfo  {journal} {Optics Letters}\ }\textbf {\bibinfo {volume}
  {32}},\ \bibinfo {pages} {14} (\bibinfo {year} {2007})}\BibitemShut {NoStop}%
\bibitem [{\citenamefont {Shain}\ \emph {et~al.}(2017)\citenamefont {Shain},
  \citenamefont {Vickers}, \citenamefont {Goldberg}, \citenamefont {Bifano},\
  and\ \citenamefont {Mertz}}]{Shain}%
  \BibitemOpen
  \bibfield  {author} {\bibinfo {author} {\bibfnamefont {W.~J.}\ \bibnamefont
  {Shain}}, \bibinfo {author} {\bibfnamefont {N.~A.}\ \bibnamefont {Vickers}},
  \bibinfo {author} {\bibfnamefont {B.~B.}\ \bibnamefont {Goldberg}}, \bibinfo
  {author} {\bibfnamefont {T.}~\bibnamefont {Bifano}},\ and\ \bibinfo {author}
  {\bibfnamefont {J.}~\bibnamefont {Mertz}},\ }\bibfield  {title} {\bibinfo
  {title} {Extended depth-of-field microscopy with a high-speed deformable
  mirror},\ }\href@noop {} {\bibfield  {journal} {\bibinfo  {journal} {Optics
  Letters}\ }\textbf {\bibinfo {volume} {42}},\ \bibinfo {pages} {995}
  (\bibinfo {year} {2017})}\BibitemShut {NoStop}%
\bibitem [{\citenamefont {Chen}\ \emph {et~al.}(2020)\citenamefont {Chen},
  \citenamefont {Chakraborty}, \citenamefont {Daetwyler}, \citenamefont
  {Manton}, \citenamefont {Dean},\ and\ \citenamefont
  {Fiolka}}]{ChenChakraborty_2020}%
  \BibitemOpen
  \bibfield  {author} {\bibinfo {author} {\bibfnamefont {B.}~\bibnamefont
  {Chen}}, \bibinfo {author} {\bibfnamefont {T.}~\bibnamefont {Chakraborty}},
  \bibinfo {author} {\bibfnamefont {S.}~\bibnamefont {Daetwyler}}, \bibinfo
  {author} {\bibfnamefont {J.~D.}\ \bibnamefont {Manton}}, \bibinfo {author}
  {\bibfnamefont {K.}~\bibnamefont {Dean}},\ and\ \bibinfo {author}
  {\bibfnamefont {R.}~\bibnamefont {Fiolka}},\ }\bibfield  {title} {\bibinfo
  {title} {Extended depth of focus multiphoton microscopy via incoherent pulse
  splitting},\ }\href@noop {} {\bibfield  {journal} {\bibinfo  {journal}
  {Biomedical Optics Express}\ }\textbf {\bibinfo {volume} {11}},\ \bibinfo
  {pages} {3830} (\bibinfo {year} {2020})}\BibitemShut {NoStop}%
\bibitem [{\citenamefont {Boothe}\ \emph {et~al.}(2017)\citenamefont {Boothe},
  \citenamefont {Hilbert}, \citenamefont {Heide}, \citenamefont {Berninger},
  \citenamefont {Huttner}, \citenamefont {Zaburdaev}, \citenamefont
  {Vastenhouw}, \citenamefont {Myers}, \citenamefont {Drechsel},\ and\
  \citenamefont {Rink}}]{Boothe_2017}%
  \BibitemOpen
  \bibfield  {author} {\bibinfo {author} {\bibfnamefont {T.}~\bibnamefont
  {Boothe}}, \bibinfo {author} {\bibfnamefont {L.}~\bibnamefont {Hilbert}},
  \bibinfo {author} {\bibfnamefont {M.}~\bibnamefont {Heide}}, \bibinfo
  {author} {\bibfnamefont {L.}~\bibnamefont {Berninger}}, \bibinfo {author}
  {\bibfnamefont {W.~B.}\ \bibnamefont {Huttner}}, \bibinfo {author}
  {\bibfnamefont {V.}~\bibnamefont {Zaburdaev}}, \bibinfo {author}
  {\bibfnamefont {N.~L.}\ \bibnamefont {Vastenhouw}}, \bibinfo {author}
  {\bibfnamefont {E.~W.}\ \bibnamefont {Myers}}, \bibinfo {author}
  {\bibfnamefont {D.~N.}\ \bibnamefont {Drechsel}},\ and\ \bibinfo {author}
  {\bibfnamefont {J.~C.}\ \bibnamefont {Rink}},\ }\bibfield  {title} {\bibinfo
  {title} {A tunable refractive index matching medium for live imaging cells,
  tissues and model organisms},\ }\href@noop {} {\bibfield  {journal} {\bibinfo
   {journal} {eLife}\ }\textbf {\bibinfo {volume} {6}},\ \bibinfo {pages}
  {e27240} (\bibinfo {year} {2017})}\BibitemShut {NoStop}%
\bibitem [{\citenamefont {Truong}\ \emph {et~al.}(2020)\citenamefont {Truong},
  \citenamefont {Holland}, \citenamefont {Madaan}, \citenamefont {Andreev},
  \citenamefont {Keomanee-Dizon}, \citenamefont {Troll}, \citenamefont {Koo},
  \citenamefont {McFall-Ngai},\ and\ \citenamefont {Fraser}}]{Truong_2020}%
  \BibitemOpen
  \bibfield  {author} {\bibinfo {author} {\bibfnamefont {T.~V.}\ \bibnamefont
  {Truong}}, \bibinfo {author} {\bibfnamefont {D.~B.}\ \bibnamefont {Holland}},
  \bibinfo {author} {\bibfnamefont {S.}~\bibnamefont {Madaan}}, \bibinfo
  {author} {\bibfnamefont {A.}~\bibnamefont {Andreev}}, \bibinfo {author}
  {\bibfnamefont {K.}~\bibnamefont {Keomanee-Dizon}}, \bibinfo {author}
  {\bibfnamefont {J.}~\bibnamefont {Troll}}, \bibinfo {author} {\bibfnamefont
  {D.~E.}\ \bibnamefont {Koo}}, \bibinfo {author} {\bibfnamefont
  {M.}~\bibnamefont {McFall-Ngai}},\ and\ \bibinfo {author} {\bibfnamefont
  {S.~E.}\ \bibnamefont {Fraser}},\ }\bibfield  {title} {\bibinfo {title}
  {High-contrast, synchronous volumetric imaging with selective volume
  illumination microscopy},\ }\href@noop {} {\bibfield  {journal} {\bibinfo
  {journal} {Communications Biology}\ }\textbf {\bibinfo {volume} {3}},\
  \bibinfo {pages} {74} (\bibinfo {year} {2020})}\BibitemShut {NoStop}%
\bibitem [{\citenamefont {Hect}(2017)}]{Hect}%
  \BibitemOpen
  \bibfield  {author} {\bibinfo {author} {\bibfnamefont {E.}~\bibnamefont
  {Hect}},\ }\href@noop {} {\emph {\bibinfo {title} {Optics}}},\ \bibinfo
  {edition} {5th}\ ed.\ (\bibinfo  {publisher} {Pearson Education},\ \bibinfo
  {year} {2017})\BibitemShut {NoStop}%
\bibitem [{\citenamefont {Voleti}\ \emph {et~al.}(2019)\citenamefont {Voleti},
  \citenamefont {Patel}, \citenamefont {Li}, \citenamefont {Campos},
  \citenamefont {Bharadwaj}, \citenamefont {H.~Yu}, \citenamefont {Casper},
  \citenamefont {Yan}, \citenamefont {Liang}, \citenamefont {Wen},
  \citenamefont {Kimura}, \citenamefont {Targoff},\ and\ \citenamefont
  {Hillman}}]{Voleti_2019}%
  \BibitemOpen
  \bibfield  {author} {\bibinfo {author} {\bibfnamefont {V.}~\bibnamefont
  {Voleti}}, \bibinfo {author} {\bibfnamefont {K.~B.}\ \bibnamefont {Patel}},
  \bibinfo {author} {\bibfnamefont {W.}~\bibnamefont {Li}}, \bibinfo {author}
  {\bibfnamefont {C.~P.}\ \bibnamefont {Campos}}, \bibinfo {author}
  {\bibfnamefont {S.}~\bibnamefont {Bharadwaj}}, \bibinfo {author}
  {\bibfnamefont {C.~F.}\ \bibnamefont {H.~Yu}}, \bibinfo {author}
  {\bibfnamefont {M.~J.}\ \bibnamefont {Casper}}, \bibinfo {author}
  {\bibfnamefont {R.~W.}\ \bibnamefont {Yan}}, \bibinfo {author} {\bibfnamefont
  {W.}~\bibnamefont {Liang}}, \bibinfo {author} {\bibfnamefont
  {C.}~\bibnamefont {Wen}}, \bibinfo {author} {\bibfnamefont {K.~D.}\
  \bibnamefont {Kimura}}, \bibinfo {author} {\bibfnamefont {K.~L.}\
  \bibnamefont {Targoff}},\ and\ \bibinfo {author} {\bibfnamefont {E.~M.~C.}\
  \bibnamefont {Hillman}},\ }\bibfield  {title} {\bibinfo {title} {Real-time
  volumetric microscopy of in vivo dynamics and large-scale samples with scape
  2.0},\ }\href@noop {} {\bibfield  {journal} {\bibinfo  {journal} {Nature
  Methods}\ }\textbf {\bibinfo {volume} {16}},\ \bibinfo {pages} {1054}
  (\bibinfo {year} {2019})}\BibitemShut {NoStop}%
\bibitem [{\citenamefont {Lee}\ \emph {et~al.}(2017)\citenamefont {Lee},
  \citenamefont {Andreev}, \citenamefont {Truong}, \citenamefont {Chen},
  \citenamefont {Hill}, \citenamefont {Oikonomou}, \citenamefont {Pham},
  \citenamefont {Hong}, \citenamefont {Tran}, \citenamefont {Glass},
  \citenamefont {Sapin}, \citenamefont {Engle}, \citenamefont {Fraser},\ and\
  \citenamefont {Prober}}]{Lee_2017}%
  \BibitemOpen
  \bibfield  {author} {\bibinfo {author} {\bibfnamefont {D.~A.}\ \bibnamefont
  {Lee}}, \bibinfo {author} {\bibfnamefont {A.}~\bibnamefont {Andreev}},
  \bibinfo {author} {\bibfnamefont {T.~V.}\ \bibnamefont {Truong}}, \bibinfo
  {author} {\bibfnamefont {A.}~\bibnamefont {Chen}}, \bibinfo {author}
  {\bibfnamefont {A.~J.}\ \bibnamefont {Hill}}, \bibinfo {author}
  {\bibfnamefont {G.}~\bibnamefont {Oikonomou}}, \bibinfo {author}
  {\bibfnamefont {U.}~\bibnamefont {Pham}}, \bibinfo {author} {\bibfnamefont
  {Y.~K.}\ \bibnamefont {Hong}}, \bibinfo {author} {\bibfnamefont
  {S.}~\bibnamefont {Tran}}, \bibinfo {author} {\bibfnamefont {L.}~\bibnamefont
  {Glass}}, \bibinfo {author} {\bibfnamefont {V.}~\bibnamefont {Sapin}},
  \bibinfo {author} {\bibfnamefont {J.}~\bibnamefont {Engle}}, \bibinfo
  {author} {\bibfnamefont {S.~E.}\ \bibnamefont {Fraser}},\ and\ \bibinfo
  {author} {\bibfnamefont {D.~A.}\ \bibnamefont {Prober}},\ }\bibfield  {title}
  {\bibinfo {title} {Genetic and neuronal regulation of sleep by neuropeptide
  vf},\ }\href@noop {} {\bibfield  {journal} {\bibinfo  {journal} {eLife}\
  }\textbf {\bibinfo {volume} {6}},\ \bibinfo {pages} {e25727} (\bibinfo {year}
  {2017})}\BibitemShut {NoStop}%
\bibitem [{\citenamefont {Leutenegger}\ \emph {et~al.}(2006)\citenamefont
  {Leutenegger}, \citenamefont {Rao}, \citenamefont {Leitgeb},\ and\
  \citenamefont {Lasser}}]{Leutenegger_2006}%
  \BibitemOpen
  \bibfield  {author} {\bibinfo {author} {\bibfnamefont {M.}~\bibnamefont
  {Leutenegger}}, \bibinfo {author} {\bibfnamefont {R.}~\bibnamefont {Rao}},
  \bibinfo {author} {\bibfnamefont {R.~A.}\ \bibnamefont {Leitgeb}},\ and\
  \bibinfo {author} {\bibfnamefont {T.}~\bibnamefont {Lasser}},\ }\bibfield
  {title} {\bibinfo {title} {Fast focus field calculations},\ }\href@noop {}
  {\bibfield  {journal} {\bibinfo  {journal} {Optics Express}\ }\textbf
  {\bibinfo {volume} {14}},\ \bibinfo {pages} {11277} (\bibinfo {year}
  {2006})}\BibitemShut {NoStop}%
\bibitem [{\citenamefont {Westerfield}(2000)}]{Westerfield}%
  \BibitemOpen
  \bibfield  {author} {\bibinfo {author} {\bibfnamefont {M.}~\bibnamefont
  {Westerfield}},\ }\href@noop {} {\emph {\bibinfo {title} {The Zebrafish
  Book.}}}\ (\bibinfo  {publisher} {University of Oregon Press},\ \bibinfo
  {year} {2000})\BibitemShut {NoStop}%
\end{thebibliography}%

\onecolumngrid
\appendix

\newpage
\section*{Supplementary Figures}

\setcounter{figure}{0}
\makeatletter
\renewcommand{\thefigure}{S\arabic{figure}}

\begin{figure}[!h]
\includegraphics[scale=0.3]{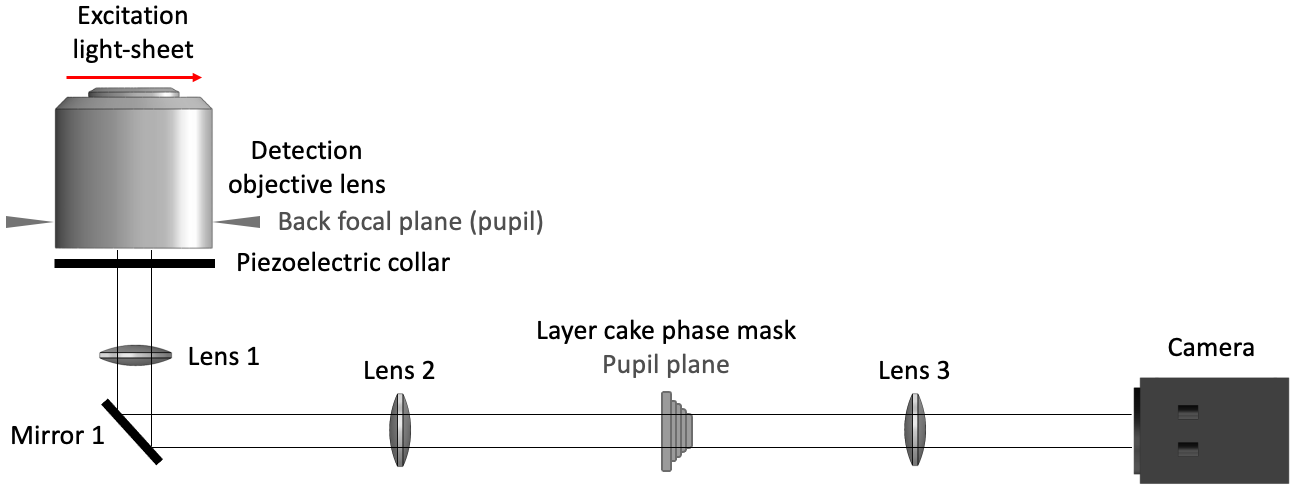}
\caption{\label{fig:ExD_optics}Simplified schematic of extended depth-of-field light-sheet microscope.\\
Simplified schematic of extended depth-of-field light-sheet microscope, illustrating the detection light path of the system. A layer-cake phase mask is positioned at a conjugate pupil plane of the detection objective to enable instantaneous depth-of-field extension to match the light-sheet thickness. See also Methods Section~\ref{sec:ExD_optics} and Fig.~\ref{fig:layercake}.}
\end{figure}

\begin{figure*}
\includegraphics[scale=0.565]{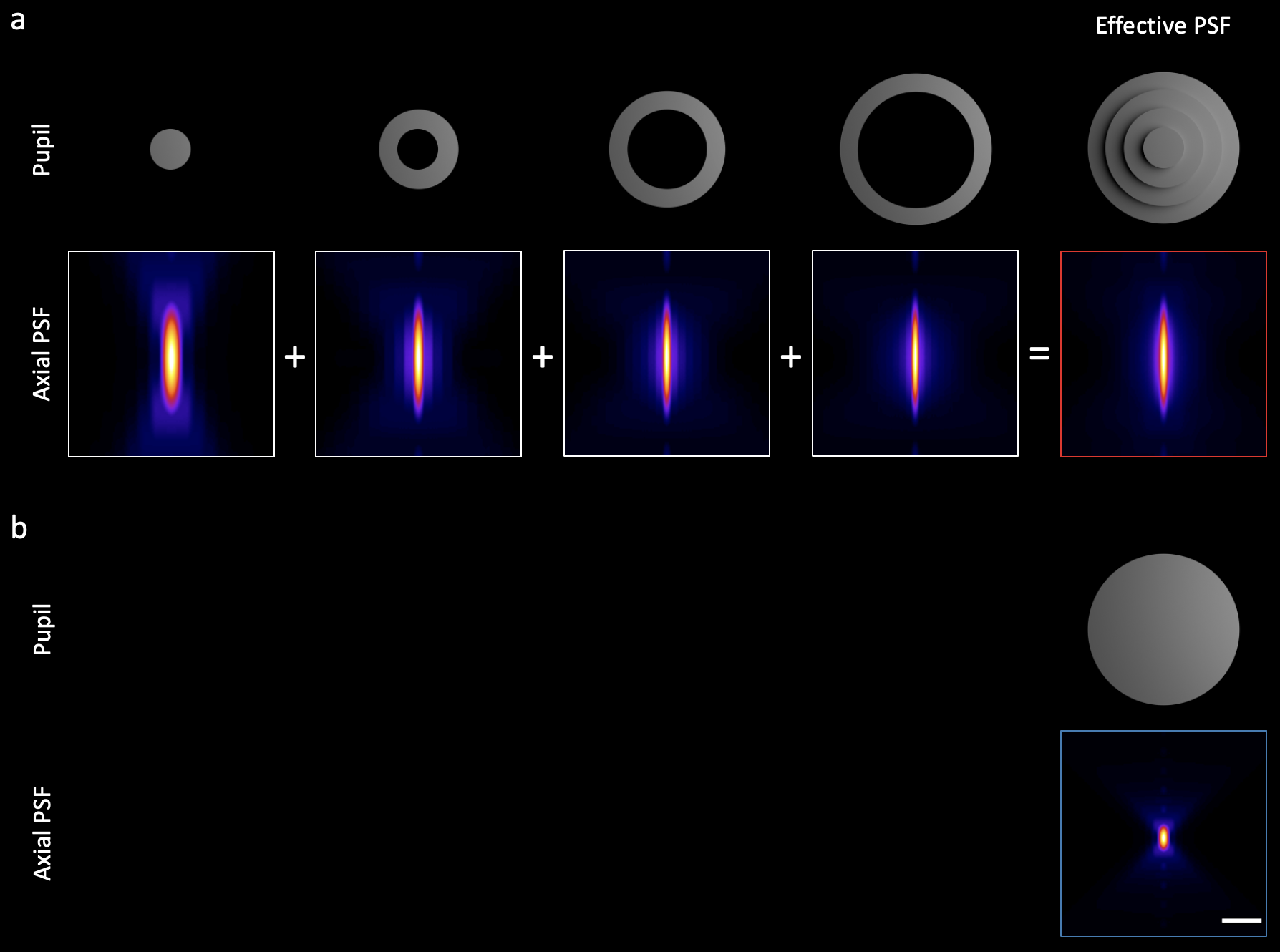}
\caption{\label{fig:layercake}Theoretical principle and simulations of extended detection point-spread functions.\\
    \textbf{(a)} A series of apertures at the pupil (top row), each of different diameter, produce the same axial extent (second row). The (left column) aperture produces a Gaussian-like focus, whereas the annular apertures (middle columns) produce Bessel-like foci. A layer-cake phase mask (right column) segments the full back-pupil of a high-NA detection objective into multiple sub-apertures (multiple concentric glass disks), each of which is $\sim\!\!400$ \si{\micro\meter} thick and thereby breaks the short coherence length of the detected light, forming mutually incoherent foci that add together to produce an axially elongated focus, without compromising light collection \cite{Abrahamsson_2006,ChenChakraborty_2020}. The amount of DOF scales linearly to the number of layers in the mask (in this case $\sim\!\!4\times$), and can thus be adjusted to the light sheet thickness.\\
    \textbf{(b)} A conventional high-NA detection objective, where different wavevectors interfere and add coherently to produce a tight focus, for comparison.\\
    Gamma was adjusted (to 0.85) in all PSF images to enhance contrast of weak features such as side lobes. Scale bar, 5 \si{\micro\meter}}
\end{figure*}

\begin{figure*}
\includegraphics[scale=0.625]{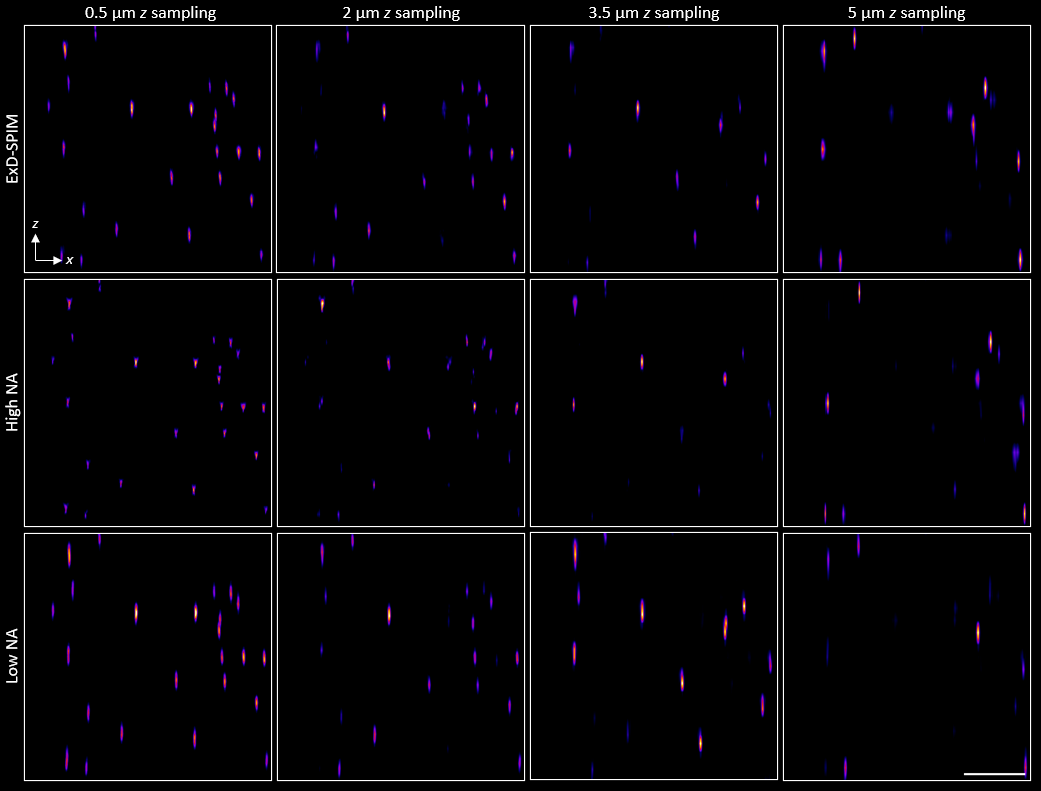}
\caption{\label{fig:undersampling}ExD-SPIM enables reduced axial sampling rates in volumetric imaging. \\
Left column: $x$-$z$ maximum intensity projection (MIP) of a 100-\si{\micro\meter}-thick sub-diffractive beads field captured with ExD-SPIM (top row), high-NA SPIM (middle row), and low-NA SPIM (bottom row). Volumetric data were acquired with 0.5 \si{\micro\meter} axial sampling for all modes.\\
Remaining columns: $x$-$z$ MIP of the same data, but re-sampled computationally with progressively larger axial sampling steps, as indicated at the top of each of the respective columns. Resampling was done by simply removing the requisite $z$-slices from the $z$-stack to produce the desired sampling condition.\\
At 3.5- and 5-\si{\micro\meter} axial sampling, both of which are coarser than the Nyquist criterion for all modalities, we see that sample information, represented by the captured beads, was clearly lost for all modalities. At 2-\si{\micro\meter} axial sampling, ExD-SPIM (and low-NA SPIM), with its extended depth of field $\gtrsim$ 5 \si{\micro\meter}, provides improved volumetric reconstruction and coverage, compared to standard high-NA SPIM (which has a depth of field $\gtrsim$ 2 \si{\micro\meter}). This not only means that more useful fluorescence signal is captured within a single image plane, but less images are required to capture the same volumetric content, which are key to reliably capturing fast dynamics over large volumes. Practically, less exposures directly translates to reduced photodamage to the specimen.\\
Display intensities set from 5\% (minimum) to 95\% (maximum) for all MIPs to compare the captured contrast range. Scale bar, 25 \si{\micro\meter}.
}
\end{figure*}

\begin{figure*}
\includegraphics[scale=0.625]{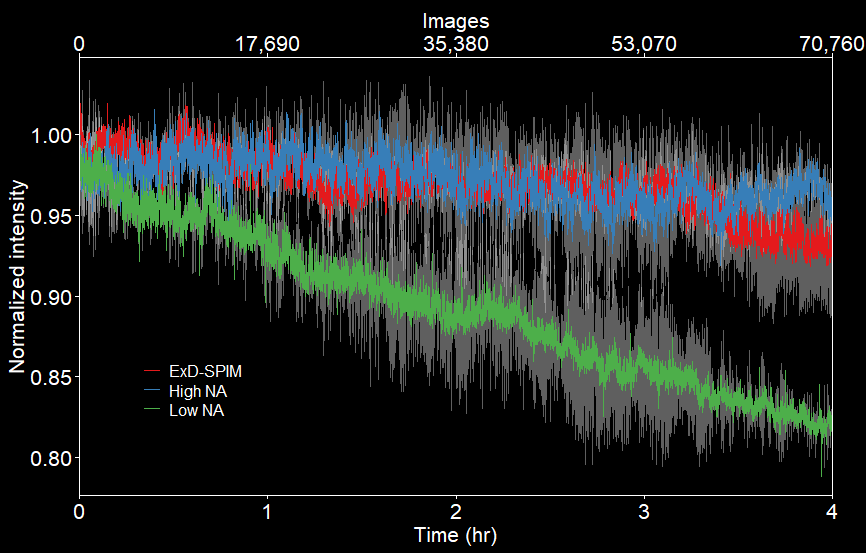}
\caption{\label{fig:photobleach}ExD-SPIM offers low photodamage \emph{in vivo}. \\
  Time-lapse images of GFP-vasculature-labeled larval zebrafish were acquired continuously over 4 hours (70,760 total images) for each modality. Plot shows the total sum intensity of the entire imaged plane as a function of time. Total intensity of ExD-SPIM and SPIM at high NA exhibit similar rates of photobleaching, decreasing by only 5\% of its initial intensity. The bleaching rate of the low-NA system is 4-fold higher, depleting signal intensity by 20\%. Shadings (grey) represent standard deviation.}
\end{figure*}

\end{document}